\documentclass[useAMS,usenatbib]{mn2e}
\usepackage{psfrag,graphicx}
\usepackage{color}
\usepackage{txfonts}
\usepackage{breqn}
\usepackage[T1]{fontenc}
\usepackage{ae,aecompl}
\title[The formation of Pop III stars]
{Disk fragmentation and the formation of population III stars}
\author[Latif et al.]
   {M. A. Latif \thanks{Corresponding author: latif@iap.fr}$^{1}$,
   D. R. G. Schleicher$^{2,3}$\\
$^1$ Institut d'Astrophysique, UMR 7095 CNRS, Universite Pierre et Marie Curie, 98bis Blvd Arago, 75014 Paris, France\\
$^2$ Institut f\"ur Astrophysik, Georg-August-Universit\"at, Friedrich-Hund-Platz 1, D-37077 G\"ottingen, Germany \\
$^3$Scuola Normale Superiore, Piazza dei Cavalieri 7, 56126 Pisa, Italy
  }
\date{}

\pagerange{\pageref{firstpage}--\pageref{lastpage}} \pubyear{2009}

\def\LaTeX{L\kern-.36em\raise.3ex\hbox{a}\kern-.15em
      T\kern-.1667em\lower.7ex\hbox{E}\kern-.125emX}

\begin{document}

\bibliographystyle{mn2e}

\label{firstpage}

\maketitle
\def\na{NewA}%
\def\aj{AJ}%
\def\araa{ARA\&A}%
\def\apj{ApJ}%
\def\apjl{ApJ}%
\def\jcap{JCAP}

\def\apjs{ApJS}%
\def\ao{Appl.~Opt.}%
\def\apss{Ap\&SS}%
\def\aap{A\&A}%
\def\aapr{A\&A~Rev.}%
\def\aaps{A\&AS}%
\def\azh{AZh}%
\def\baas{BAAS}%
\def\jrasc{JRASC}%
\def\memras{MmRAS}%
\def\mnras{MNRAS}%
\def\pra{Phys.~Rev.~A}%
\def\prb{Phys.~Rev.~B}%
\def\prc{Phys.~Rev.~C}%
\def\prd{Phys.~Rev.~D}%
\def\pre{Phys.~Rev.~E}%
\def\prl{Phys.~Rev.~Lett.}%
\def\pasp{PASP}%
\def\pasj{PASJ}%
\def\qjras{QJRAS}%
\def\skytel{S\&T}%
\def\solphys{Sol.~Phys.}%

\def\sovast{Soviet~Ast.}%
\def\ssr{Space~Sci.~Rev.}%
\def\zap{ZAp}%
\def\nat{Nature}%
\def\iaucirc{IAU~Circ.}%
\def\aplett{Astrophys.~Lett.}%
\def\apspr{Astrophys.~Space~Phys.~Res.}%
\def\bain{Bull.~Astron.~Inst.~Netherlands}%
\def\fcp{Fund.~Cosmic~Phys.}%
\def\gca{Geochim.~Cosmochim.~Acta}%
\def\grl{Geophys.~Res.~Lett.}%
\def\jcp{J.~Chem.~Phys.}%
\def\jgr{J.~Geophys.~Res.}%
\def\jqsrt{J.~Quant.~Spec.~Radiat.~Transf.}%
\def\memsai{Mem.~Soc.~Astron.~Italiana}%
\def\nphysa{Nucl.~Phys.~A}%
\def\physrep{Phys.~Rep.}%
\def\physscr{Phys.~Scr}%
\def\planss{Planet.~Space~Sci.}%
\def\procspie{Proc.~SPIE}%

%


\begin{abstract}
{Our understanding of population III star formation is still in its infancy. They are formed in dark matter minihalos of $\rm 10^5-10^6~M_{\odot}$ at $z=20-30$. Recent high resolution cosmological simulations show that a protostellar disk forms as a consequence of gravitational collapse and fragments into multiple clumps. However, it is not entirely clear if these clumps will be able to survive to form multiple stars as simulations are unable to follow the disk evolution for longer times. In this study, we employ a simple analytical model to derive the properties of marginally stable steady-state disks. Our results show that the stability of the disk depends on the critical value of the viscous parameter $\alpha$. For $\alpha_{crit} = 1$, the disk is stable for an accretion rate of $\rm \leq 10^{-3}~M_{\odot}/yr$ and becomes unstable at radii about $\rm \geq 100~AU$ in the presence of an accretion rate  of $\rm 10^{-2}~M_{\odot}/yr$. For $0.06 < \alpha_{crit} < 1$, the disk can be unstable for both accretion rates. The comparison of the migration and the Kelvin-Helmholtz time scales shows that clumps are expected to migrate inward before reaching the main sequence. Furthermore, in the presence of a massive central star the clumps within the central 1 AU will be tidally disrupted. We also find that UV feedback from the central star is unable to disrupt the disk, and that photo-evaporation becomes important only once the accretion rate has dropped to $\rm 2~ \times~ 10^{-4}~M_{\odot}/yr$. As a result, the central star may reach a mass of 100 $\rm M_{\odot}$ or even higher.
}
\end{abstract}



\section{Introduction}
The first generation of stars, known as the population III stars ushered the cosmos out of the dark ages and brought the first light. They reionized the universe via their stellar radiation and enriched the intergalactic medium with metals through supernova explosions. They further influenced the subsequent structure formation through their chemical, mechanical and radiative feedback \citep{2004PASP..116..103B,2005SSRv..116..625C,Schleicher08,2014arXiv1410.3482G}. Hence, their comprehension is a matter of prime astrophysical interest and intrinsically linked to the understanding of key epochs in the universe. 

According to the hierarchical paradigm of structure formation, population III (Pop III) stars are assembled in primordial dark matter halos of $\rm 10^5-10^6~M_{\odot}$ at $z=20-30$. The gas falls into the dark matter potential and gets heated up to the virial temperature of the halo (i.e. $\rm \sim$ 2000 K). In the absence of metals, $\rm H_2$ is the only coolant which can bring the temperature down to a few hundred K and induce star formation. The trace amount of molecular hydrogen formed via gas-phase reactions gets boosted during the process of virialization and triggers the collapse in minihalos. At high densities ($\rm 10^8~cm^{-3}$), the three-body reactions turn the gas into a fully molecular state and finally the gas becomes optically thick to molecular hydrogen cooling above $\rm 10^{18}~cm^{-3}$. At this stage the thermal evolution becomes adiabatic and the birth of the Pop III stars takes place \citep{Omukai2000,Yoshida2006}.

The first numerical simulations performed to study the formation of Pop III stars showed that they had masses of about a few hundred solar and were living isolatory lives \citep{Abel2000,Abel2002,Bromm2002,Yoshida2006}. However, cosmological simulations performed during the past few years have improved the modeling of chemical processes and employed higher resolution below the AU scales \citep{Turk09,Clark11,Greif12,Bovino13,LatifPopIII13,Hartwig2014}. These studies show that a protostellar disk forms as a consequence of gravitational collapse and becomes unstable in the presence of higher accretion rates of about $\rm 10^{-2}-10^{-3}~M_{\odot}/yr$. Consequently, multiple stars are expected to form per halo. These works have tremendously enhanced our understanding of primordial star formation. They are however unable to follow the collapse for longer time scales due to the computational constraints. Hence, it is not clear if these secondary fragments can survive to form stars or get merged with the central star. Recently, further attempts have been made to include the stellar UV feedback from the central star \citep{Hosokawa14,Stacy2012,Hirano2014,Susa14} and to derive the initial mass function. These studies either include an approximate treatment of stellar feedback or allow only a single star per halo. 

Motivated by the above mentioned results, we derive the properties of a marginally unstable protostellar disk and the fate of the clumps formed via gravitational instabilities. To achieve this goal, we employ an analytical model to compute the expected masses of the clumps and the disk under various conditions. We further show how the chemical processes influence the stability of the disk and explore the potential impact of the stellar feedback. This study provides insight into the disk fragmentation and ultimately the masses of the first stars.

This paper is organized in the following way. In the 2nd section, we describe in detail our analytical model and assumptions behind it. We present our results in section 3. In subsection 3.1, we discuss the disk properties for the disk mass dominated case where most of mass lies inside the disk at initial stage. In subsection 3.2, we discuss the disk fragmentation in the presence of central star of various masses. In section 3.3, we discuss the dependence of our results on the critical value of the viscous parameter $\alpha$. In subsection 3.4, we discuss the potential impact of UV feedback from the central star. In section 4, we summarize our findings and discuss the caveats in our model.

\section{Model}
Our model is mainly based on \cite{Inayoshi2014}. We consider the properties of a protostellar disk forming in a minihalo of $\rm 10^5-10^6~M_{\odot}$ at $z=25$. The disk is metal-free, composed of a primordial gas composition, fully molecular and predominantly cooled by the molecular hydrogen. The typical accretion rates onto such a disk are expected to be in the range of $\rm 10^{-2}-10^{-4}~M_{\odot}/yr$ as found from numerical simulations \citep{LatifPopIII13}. The main difference between our current study and \cite{Inayoshi2014} is that they studied the fragmentation of the disk forming in the massive primordial halo of $\rm \sim 10^8~M_{\odot}$ mainly cooled by atomic line cooling. Particularly, they investigated the impact of $\rm H^-$ cooling occurring at high densities which induces fragmentation in the inner part of the disk, while we focus on the disk forming around Pop III star in minihalo cooled by molecular hydrogen.  

To study the fragmentation of a self-gravitating protostellar accretion disk, we assume that it is in a steady state condition. The stability of the disk can be measured from the Toomre parameter Q which is defined as:
\begin{equation}
 Q = \frac{c_{s} \Omega}{\pi G \Sigma}
\label{eq1}
\end{equation}
$\Omega$ is the orbital frequency, $c_s$ is the sound speed, G is the gravitational constant and $\Sigma$ the surface density of the disk given as:
\begin{equation}
 \Sigma = \frac{\dot{M}_{tot}}{3 \pi \nu}
\label{eq2}
\end{equation}
$\dot{M}_{\rm tot}$ is the mass accretion rate. We consider here the cases with $\dot{M}_{\rm tot}= 10^{-2}~ \& ~10^{-3}$M$_{\odot}$/yr. $\nu$ is the viscosity in the disk arising from gravitational torques. The disk is presumed to be marginally stable with Toomre Q = 1. The scale-height of the disk is computed assuming hydrostatic equilibrium in the vertical direction 
\begin{equation}
 H = \frac{c_s}{\Omega}
\label{eq3}
\end{equation}
and the particle density is defined as
\begin{equation}
 n = \frac{\Sigma}{4m_pH}
\label{eq4}
\end{equation}
where $m_p$ is the proton mass. The thermal state of the disk can be obtained by assuming thermal equilibrium, i.e. the  heating rate ($Q_{+}$) is equivalent to cooling rate ($Q_{-}$). The main source of heating in the disk is the viscous heating by turbulence and shocks which can be computed as \citep{2007NCimR..30..293L}:
\begin{equation}
 Q_+ = \frac{9}{4} \nu \Sigma \Omega^2
\label{eq5}
\end{equation}
and cooling rate is given by:
\begin{equation}
 Q_- = 2 H \Lambda_{H_{2}/CIE}
\label{eq6}
\end{equation}
here $\Lambda_{H_{2}/CIE}$ is the net cooling rate in units of $\rm erg/cm^3/s$ and $H$ is the height of the disk. The $\rm H_2$ line cooling becomes optically thick at densities above $\rm 10^8~cm^{-3}$ and is estimated in the following way: 
\begin{equation}
\Lambda_{H_2} = \Lambda_{H2,thin} f_{thick},
\end{equation}
where $\Lambda_{H_2, thin}$ is the optically thin $\rm H_2$ line cooling rate and $f_{thick}$ is the factor which takes into account the opacity corrections. The expression for $f_{thick}$ is taken from \cite{Ripamonti2004} and is given as $min \left[1,\left(n_{H_2}/(8 \times 10^9~cm^{-3})\right)^{-0.45}\right]$.
By solving the thermal balance with respect to $\Omega$ assuming gas in the disk is fully molecular, we get the following relation:
\begin{equation}
\Omega^2 = \frac{24 \pi c_s}{9 \dot{M}_{tot}} \frac{(8 \times 10^9)^{0.45}} {\sqrt{4 \pi m_p G}} \Lambda_{H_2}  
\label{eq9}
\end{equation} 
For $\rm H_{2,thin}$ cooling, we use the cooling function for high density limit given in \cite{1998A&A...335..403G} originally computed by \cite{Hollenbach79}. We take into account the cooling due to the CIE of the molecular hydrogen which becomes dominant coolant at densities above $\rm 10^{13}~cm^{-3}$, often referred as $\rm H_2$ continuum cooling \citep{Omukai2000,Ripamonti2004}. The importance of CIE cooling has  been investigated in detail in numerical simulations \citep{Yoshida2008,Hirano2013}. They found that it influences the cloud morphology and may even induce fragmentation. For the regime dominated by CIE cooling, we get the following expression for $\Omega$ by solving the thermal balance:
\begin{equation}
\Omega = \frac{9 \dot{M}_{tot}}{24 \pi c_s} \frac{(4 \pi m_p G)^{2}}{\Lambda_{CIE}}  
\label{eq91}
\end{equation} 
$\Lambda_{CIE}$ is the optically thin CIE cooling rate  and we use the fit given in equation 17 of \cite{2014MNRAS.439.2386G}.

Our treatment for the CIE cooling is relatively simple and does not include radiative transfer effects. For a spherical collapse, at densities above $\rm 10^{16}~cm^{-3}$ the gas starts to become optically thick to the CIE cooling. However, for the disk geometry, an optically thin approximation may be valid even up to higher densities. We therefore expect at most a minor correction due to the approximation employed here, which has however no influence on our main results. Additional contributions to the thermal evolution may come from the chemical heating and cooling due to the formation and dissociation of $\rm H_2$, which is not included in our model. For a stationary disk, we assume here that also the chemical abundances have reached their equilibrium value, and that no further chemical heating or cooling occurs. However, as simulations show \citep{Omukai2000,LatifPopIII13,Bovino13}, these processes can be relevant at earlier stages or at the transition point between different regimes.

For a viscous disk, we use  the standard description for $\alpha$ by \cite{Shakura73} and is given as:
\begin{equation}
 \nu = \alpha c_s H ~. 
\label{eq7}
\end{equation} 
The properties of the fragmenting disk are computed based on above equations. The viscous parameter $\alpha$ can be computed by combing equations \ref{eq1}, \ref{eq2}, \ref{eq3} and \ref{eq7} as
\begin{equation}
\alpha = \frac{\dot{M}_{tot} G}{ 3 c_s^3}~. 
\label{eq12}
\end{equation} 
The disk becomes unstable for $\alpha > \alpha_{crit}$ \citep{Gammie01}. The initial masses of clumps can be estimated as follows:
\begin{equation}
M_{c,i} = \Sigma H^2  
\label{eq13}
\end{equation} 
The clumps in the disk  grow via accretion. The maximum clump mass so-called the gap-opening mass is computed by using the relation given in \cite{2007MNRAS.374..515L}:
\begin{equation}
M_{c} =  M_{c,i} \left[12 \pi \frac{\alpha}{0.3}\right]^{1/2} \left(\frac{R}{H}\right)^{1/2}
\label{eq14}
\end{equation} 
here R is the radius of the disk. The Kelvin-Helmholtz (KH) contraction time scale estimated in \cite{Inayoshi2014} is equivalent to the mass accretion time scale and allows to estimate the duration of the adiabatic phase of protostellar evolution. Here, we instead consider the protostellar contraction phase, which is more relevant in the presence of lower accretion rates, it is computed as \citep{Hosokawa12}: 
\begin{equation}
t_{KH} = \frac{G M_*^2}{R_* L_*} 
\label{eq15}
\end{equation}
here  $M_*$ is the mass, $R_*$ is the radius and $L_*$ is the luminosity of the protostar. We use equation 4 of \cite{Hosokawa12} to compute the luminosity and their Fig 5 to estimate the radius of the star. The mass accretion rate onto the clumps and is defined as
\begin{equation}
 \dot{M}_c =  \frac{3}{2} \Sigma \Omega \left( R_H \right)^2
\label{eq16}
\end{equation}
where R$_H$ is the Hill radius given as
\begin{equation}
 R_H =  R \left( M_c /30 M_* \right)^{1/3}
\label{eq17}
\end{equation}
$M_*$ is the mass of central star and is taken as 1 $\rm M_{\odot}$. The migration time is considered to be equivalent to the viscous time scale and is given as \citep{Lin86}:
\begin{equation}
t_{mig} =  \frac{1}{3 \pi \alpha} \left( \frac{R}{H} \right)^{2} \frac{2 \pi}{\Omega} .
\label{eq18}
\end{equation}
This approximation is valid for the type-II migration in the theory of planet-disk interactions. The type-I migration time scale is valid for low-mass planets when the perturbations induced by them are in the linear regime. It is even shorter than the type-II migration time scale. So, our choice of the migration time scale is rather conservative. We expect this time scale to be shorter for the clumpy disks. The clumps can be further destroyed by the tidal forces exerted by central star. To estimate this effect, we compute the Roche limit which is given as
\begin{equation}
R_{Roche} =  1.26 \times R_c \left( M_* / M_c \right)^{1/3}
\label{eq19}
\end{equation}
where $R_{\rm c}$, $M_{\rm c}$ are the radius and mass of the secondary clump and $M_{\rm *}$ is the mass of the central star. 

\subsection{Disk mass dominated case}
In this subsection, we consider the early phase of the disk evolution where most of the mass lies within the disk. We study the disk structure and explore the fragmentation properties of the disk. In this case, the relation between the orbital frequency and radius obeys the following relation:
\begin{equation}
 \Omega = \frac{2 c_s}{R}  
\label{eqn}
\end{equation} 
The surface density of the disk can be written as a function of orbital frequency:
\begin{equation}
 \Sigma = \frac{\Omega c_s}{\pi G}  
\label{eq10}
\end{equation} 
and the density inside the disk is
\begin{equation}
n = \frac{\Omega^2}{4 \pi m_p G}  
\label{eq11}.
\end{equation}
The rest of the quantities can be expressed in terms of radius by employing the relation given in equation \ref{eqn}.

\subsection{Star mass dominated case}
In the previous subsection, we assumed that the mass of the central star is negligible and that most of the mass lies within the disk. However, once a star is formed it will grow via rapid accretion and may change the disk structure. In this case the mass of the central star exceeds the disk mass and modifies the relation between the orbital frequency and the disk radius. Assuming Kepler rotation, we have:
\begin{equation}
R =  \left( G M_* / \Omega^2 \right)^{1/3} ,
\label{eq19}
\end{equation}
where $M_*$ is the mass of the central star. The surface density of the disk can be computed as follows:
\begin{equation}
 \Sigma = \frac{c_s}{\pi G}  \sqrt{\frac{GM_{*}}{R^3}} 
\label{eq20}
\end{equation} 
and the density inside the disk is
\begin{equation}
n = \frac{\Sigma}{4 \pi m_p c_s}  \sqrt{\frac{GM_{*}}{R^3}} 
\label{eq21}.
\end{equation}
Similarly, the rest of the properties can be computed by replacing  $\Omega= \sqrt{\frac{GM_{*}}{R^3}}$ in the equations given above.

\begin{figure*}
\hspace{-6.0cm}
\centering
\begin{tabular}{c c}
\begin{minipage}{6cm}
\vspace{-0.6cm}
\includegraphics[scale=0.7]{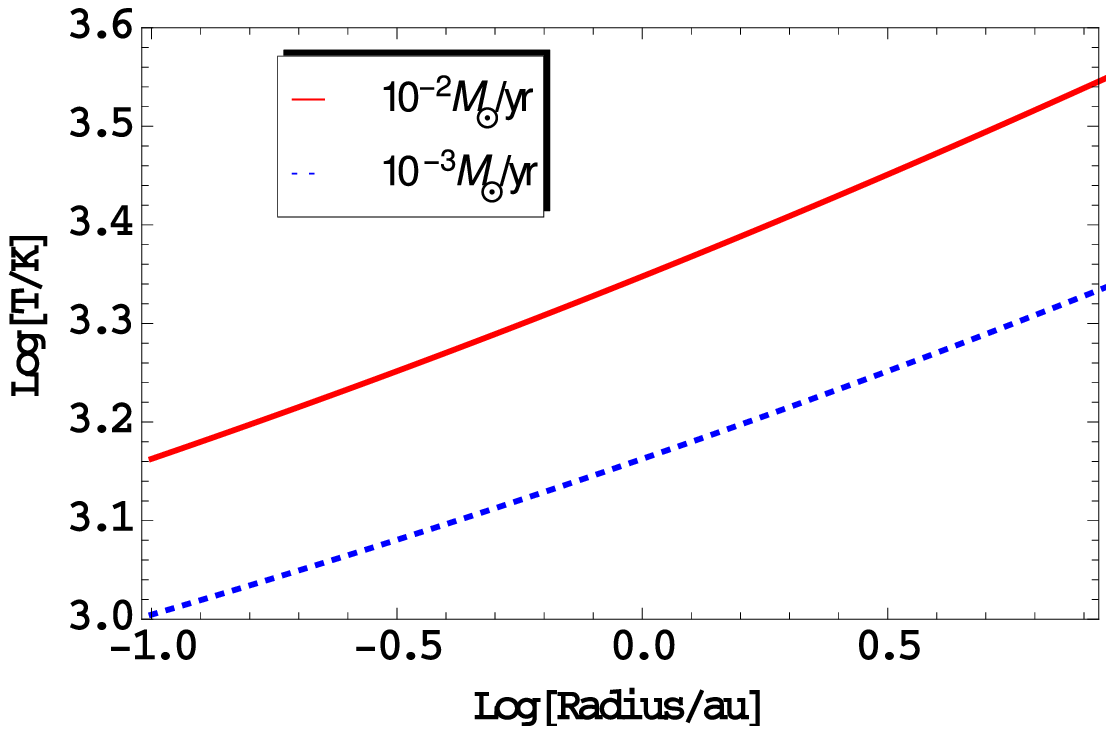}
\end{minipage}&
\begin{minipage}{6cm}
\hspace{2cm}
\includegraphics[scale=0.7]{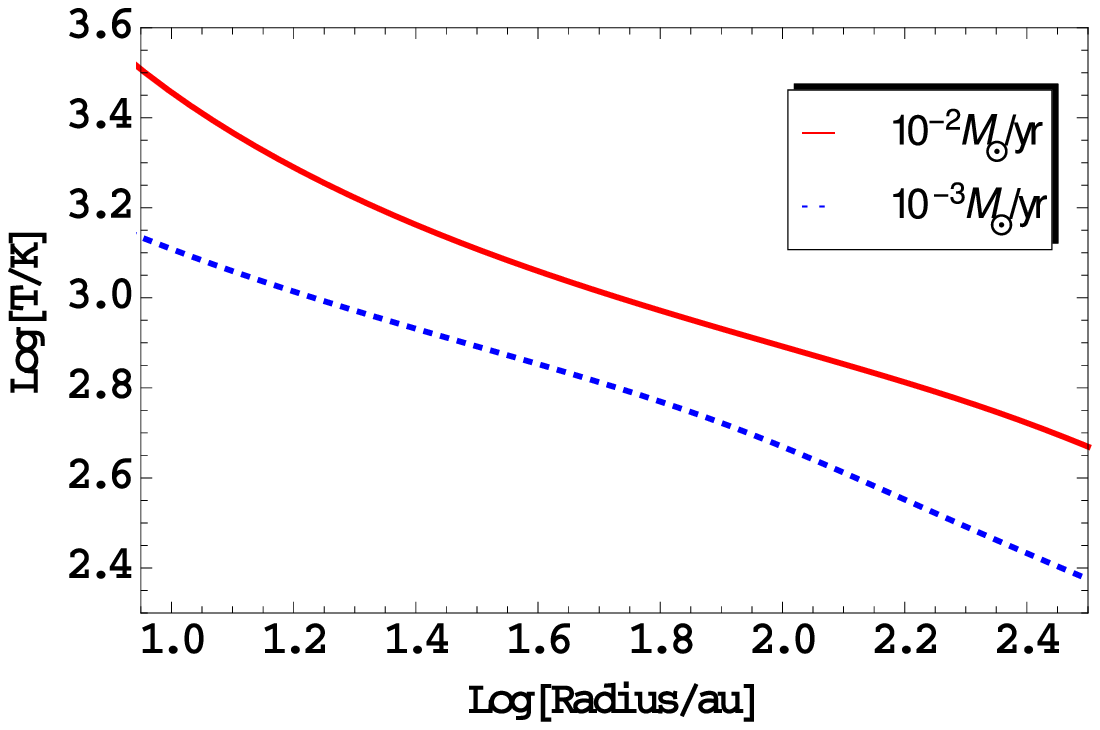}
\end{minipage} \\
\begin{minipage}{6cm}
\vspace{-0.6cm}
\includegraphics[scale=0.7]{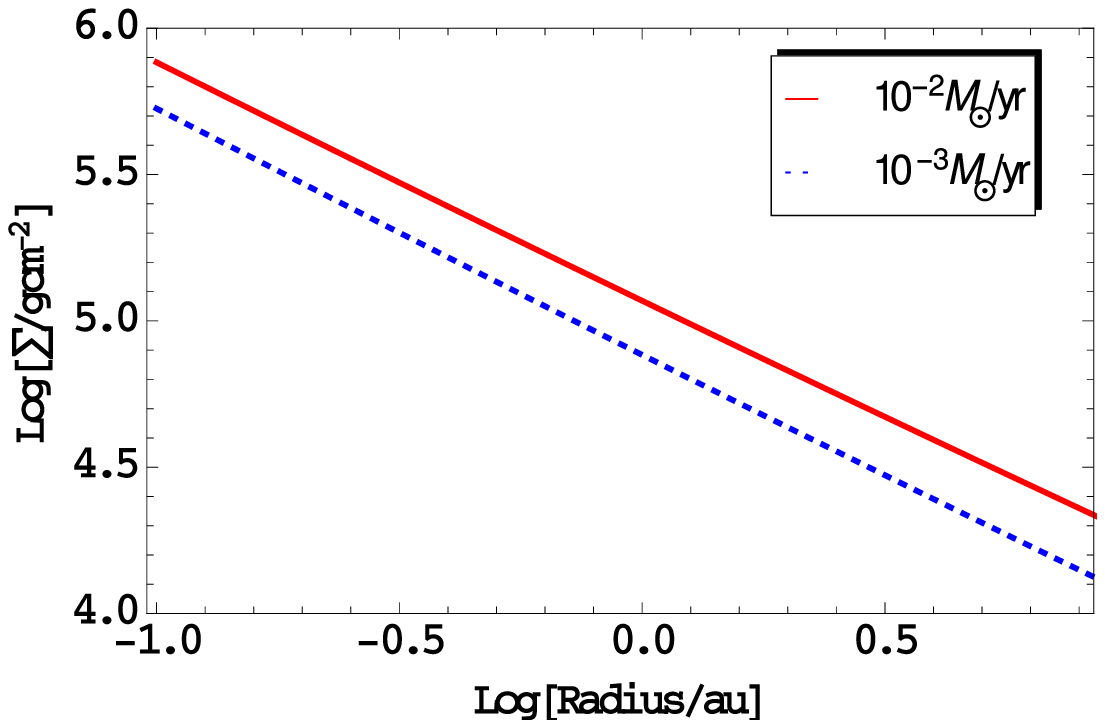}
\end{minipage}&
\begin{minipage}{6cm}
 \hspace{2cm}
\includegraphics[scale=0.7]{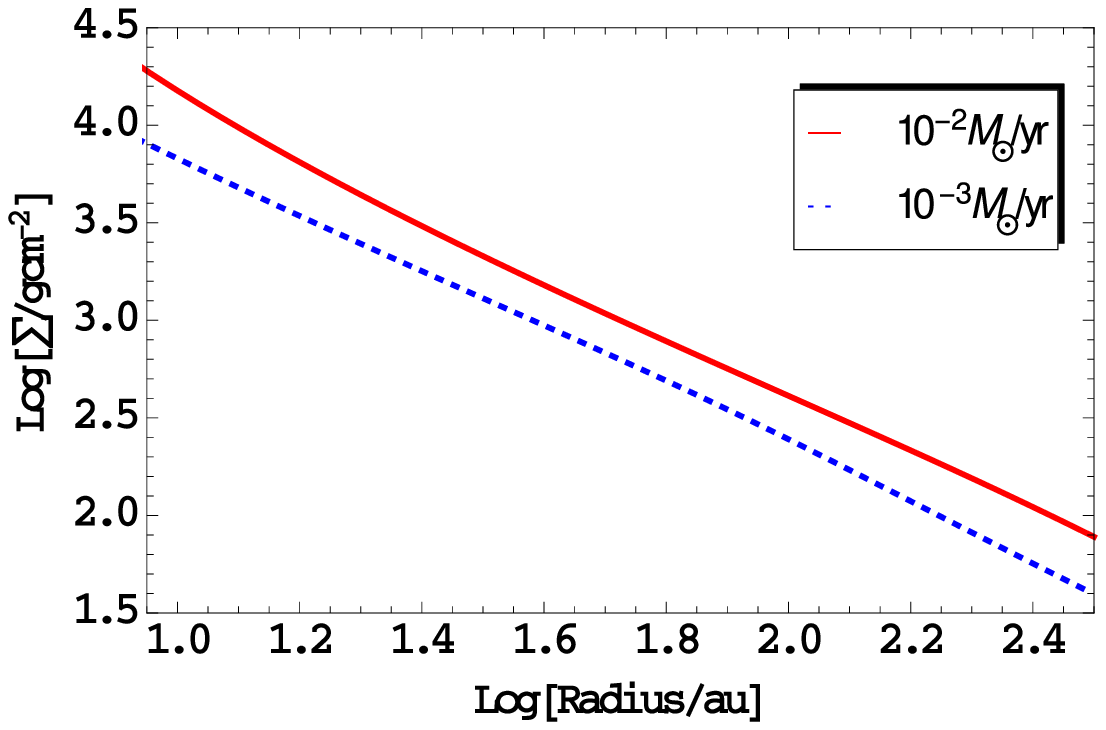}
\end{minipage} \\
\begin{minipage}{6cm}
\vspace{-0.6cm}
\includegraphics[scale=0.7]{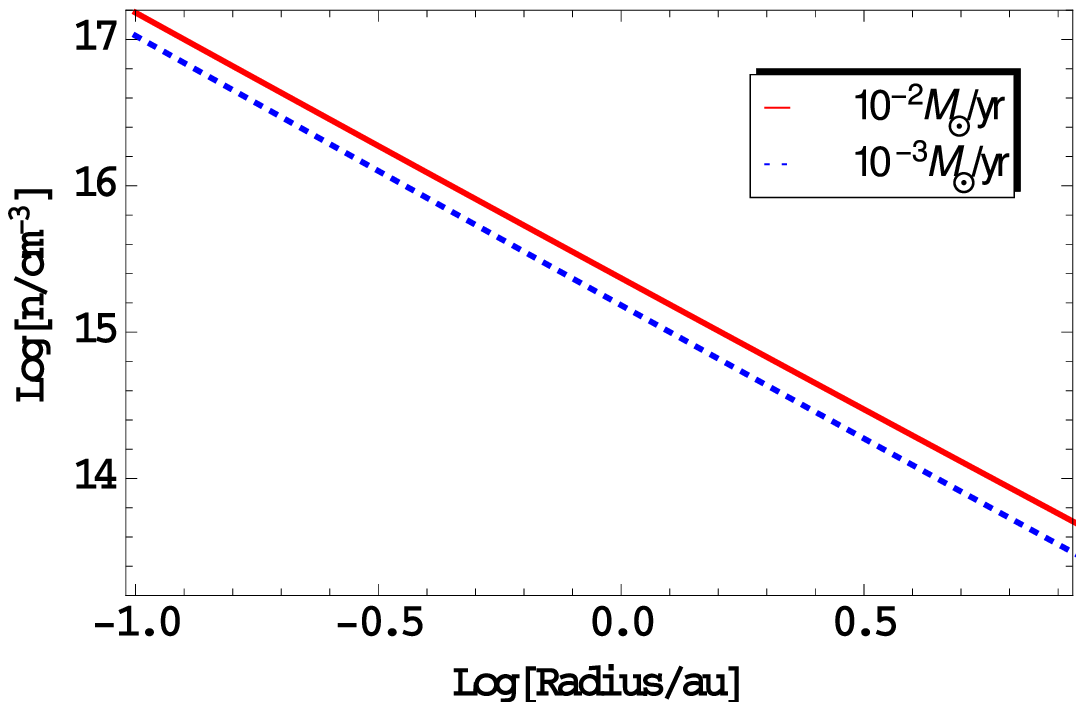}
\end{minipage}&
\begin{minipage}{6cm}
 \hspace{2cm}
\includegraphics[scale=0.7]{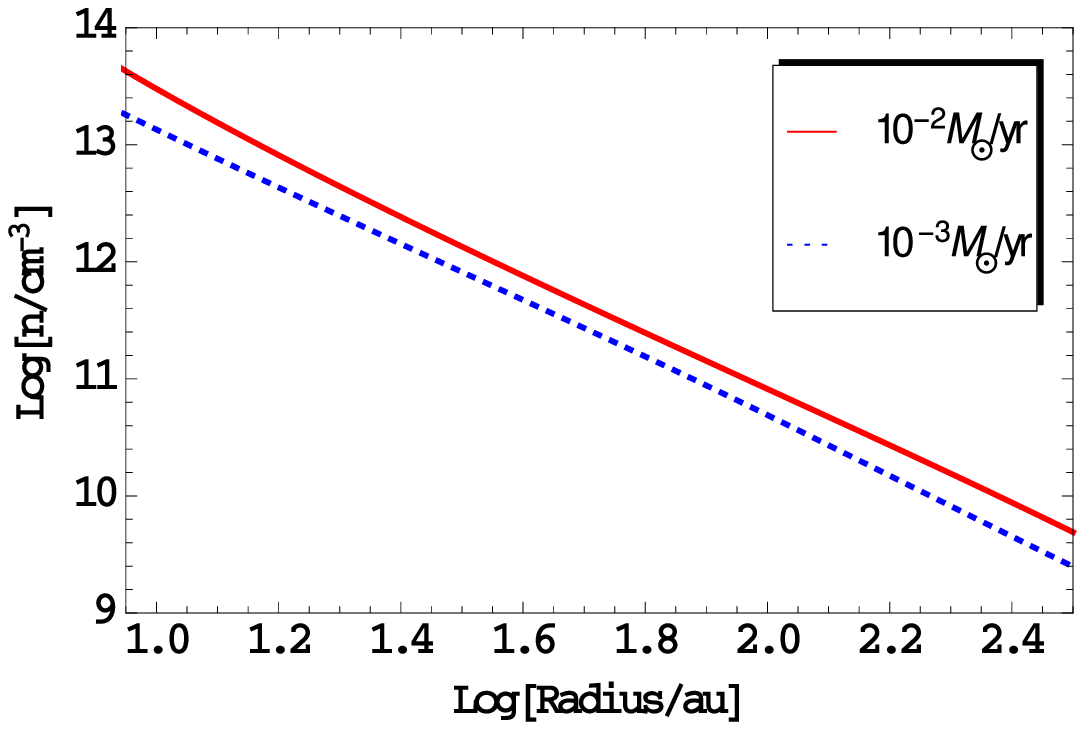}
\end{minipage} \\
\begin{minipage}{6cm}
\vspace{-0.6cm}
\includegraphics[scale=0.7]{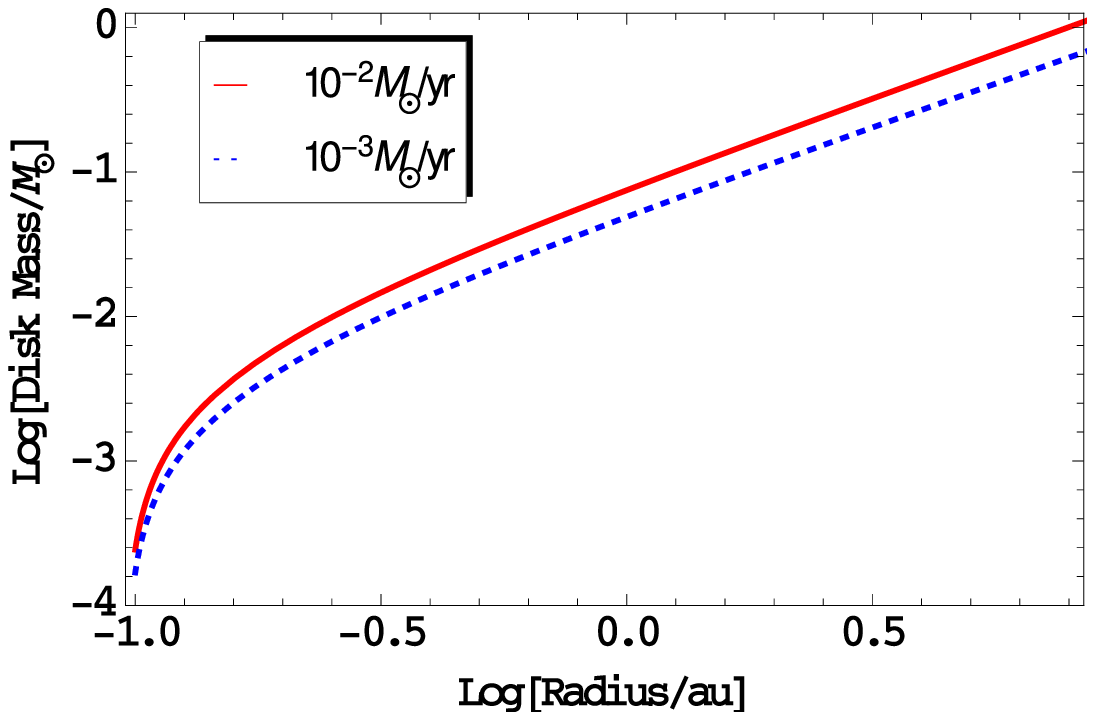}
\end{minipage}&
\begin{minipage}{6cm}
 \hspace{2cm}
\includegraphics[scale=0.7]{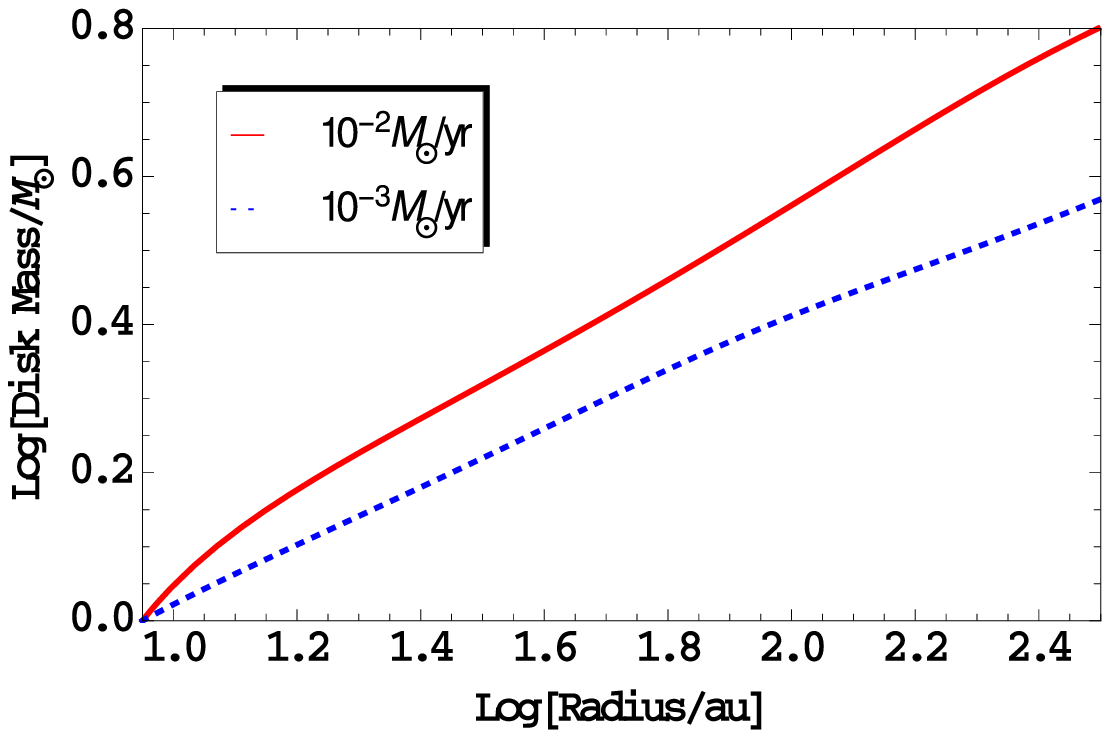}
\end{minipage} 
\end{tabular}
\caption{The properties of the disk are plotted against the radius for the disk mass dominated case. The red and blue lines represent accretion rates of $\rm 10^{-2}~M_{\odot}$/yr and $\rm 10^{-3}~M_{\odot}$/yr respectively. The left panel shows the CIE cooling regime while the right panel depicts $\rm H_2$ cooling regime.  The temperature, the density, the surface density and the enclosed mass in the disk are shown here.}
\label{fig}
\end{figure*}

\begin{figure*}
\hspace{-6.0cm}
\centering
\begin{tabular}{c c}
\begin{minipage}{6cm}
\vspace{-0.6cm}
\includegraphics[scale=0.7]{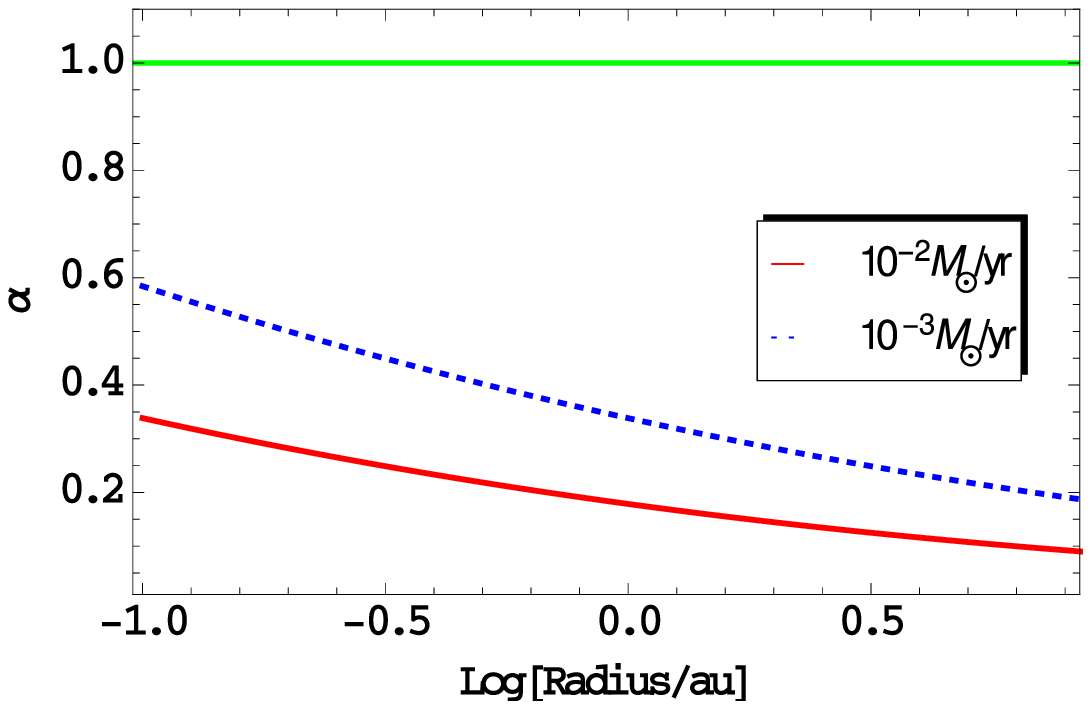}
\end{minipage}&
\begin{minipage}{6cm}
 \hspace{2.6cm}
\includegraphics[scale=0.7]{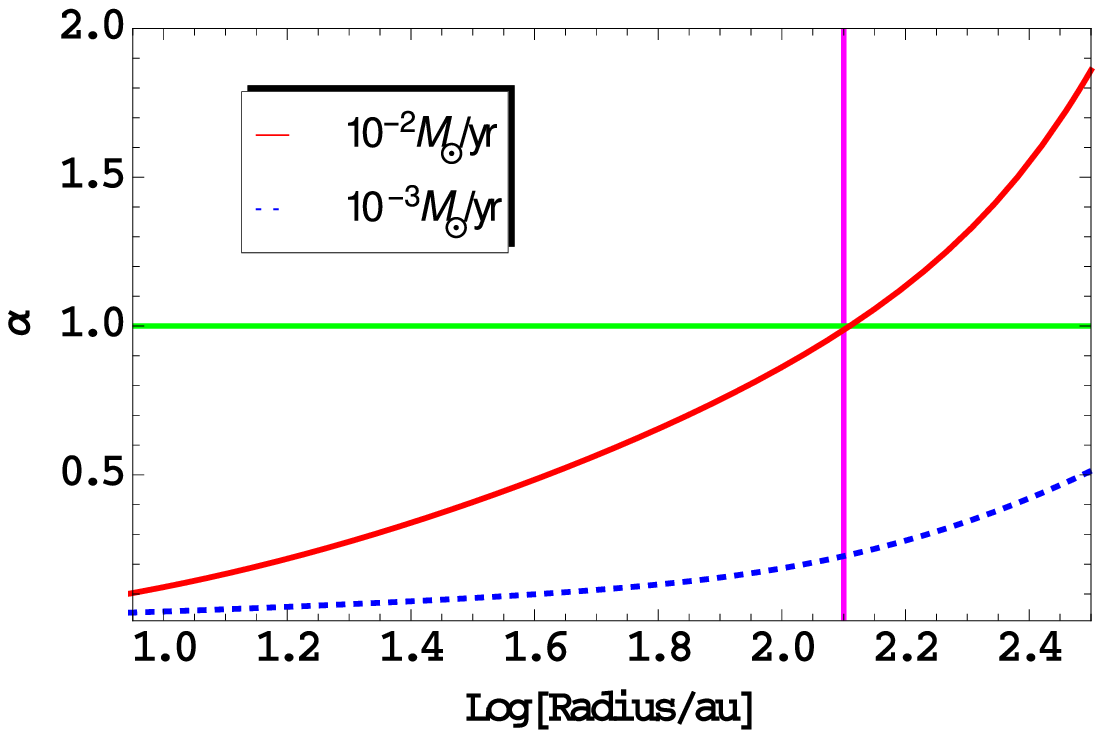}
\end{minipage} \\
\begin{minipage}{6cm}
\vspace{-0.6cm}
\includegraphics[scale=0.7]{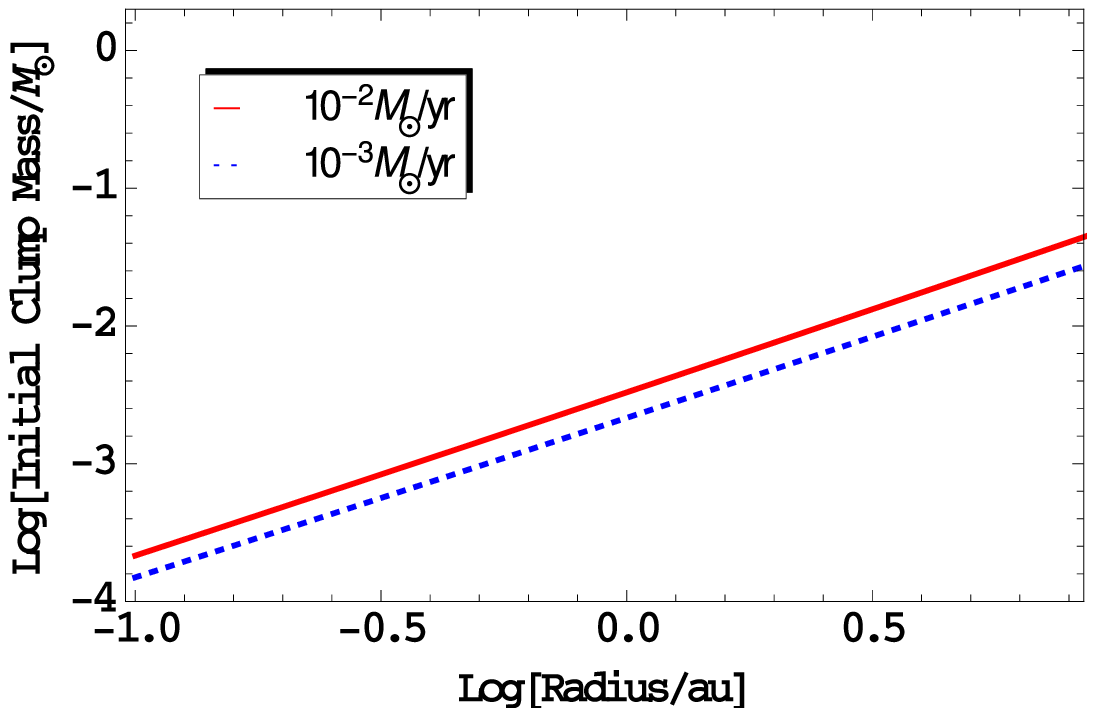}
\end{minipage}&
\begin{minipage}{6cm}
 \hspace{2cm}
 \includegraphics[scale=0.7]{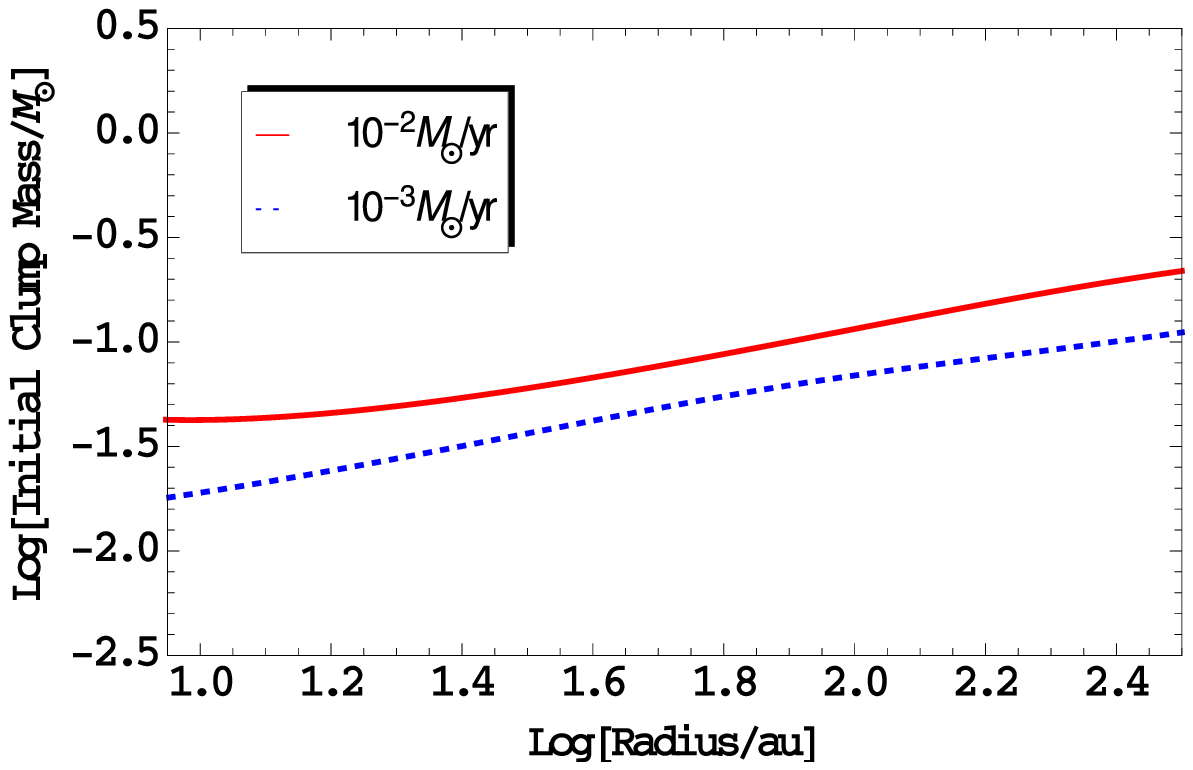}
\end{minipage} \\
\begin{minipage}{6cm}
\vspace{-0.6cm}
\includegraphics[scale=0.7]{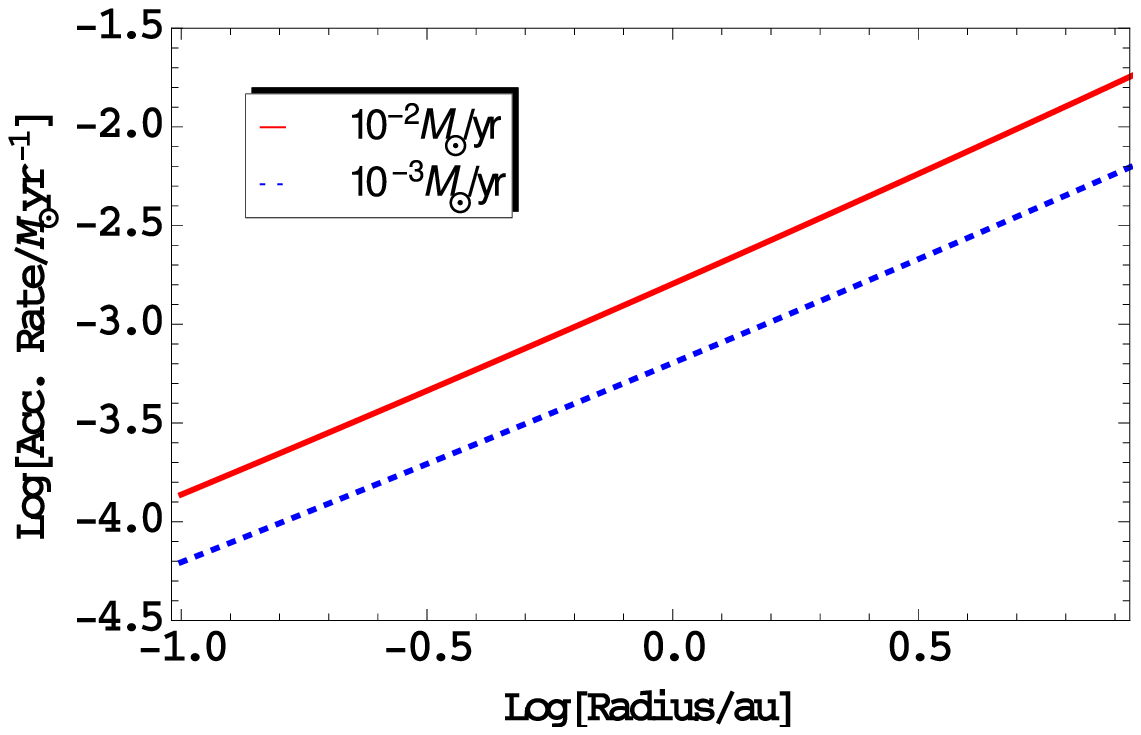}
\end{minipage} &
\begin{minipage}{6cm}
 \hspace{2.4cm}
\includegraphics[scale=0.7]{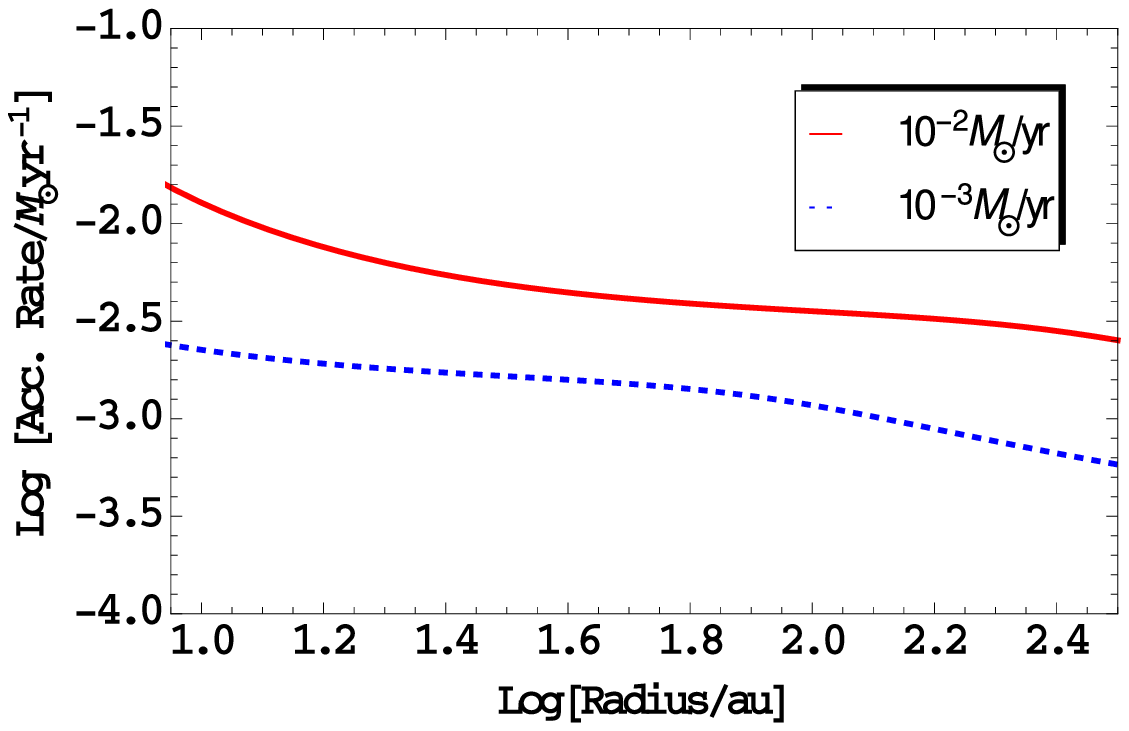}
\end{minipage} \\
\begin{minipage}{6cm}
\vspace{-0.6cm}
\includegraphics[scale=0.7]{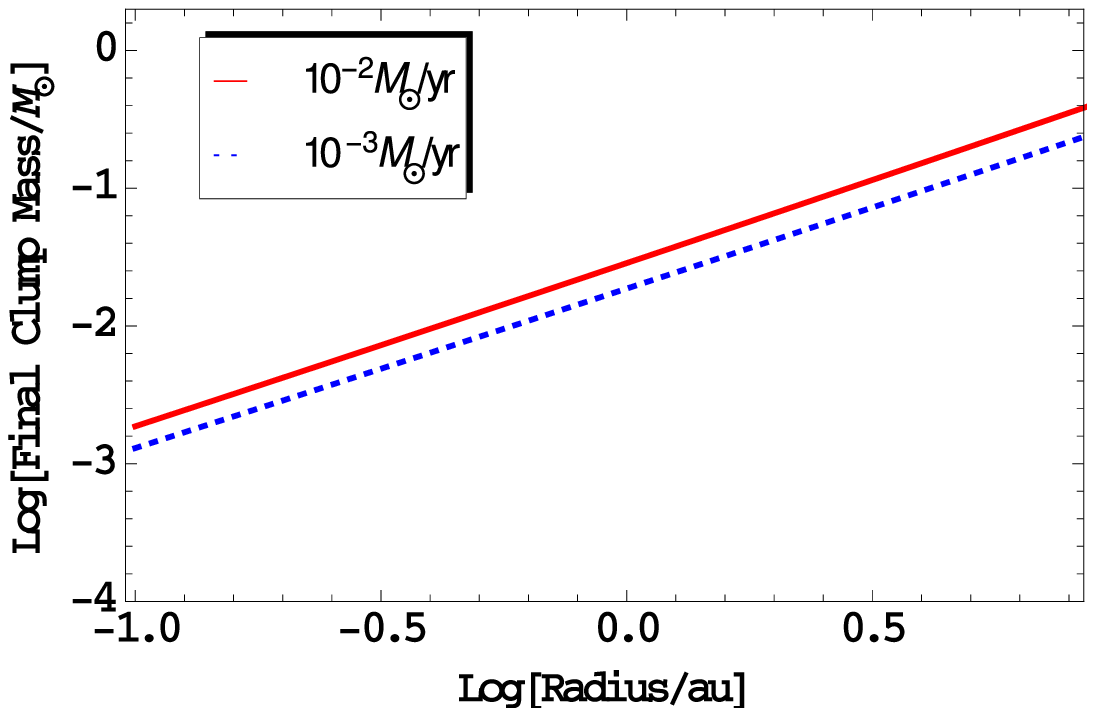}
\end{minipage} &
\begin{minipage}{6cm}
 \hspace{2.4cm}
\includegraphics[scale=0.7]{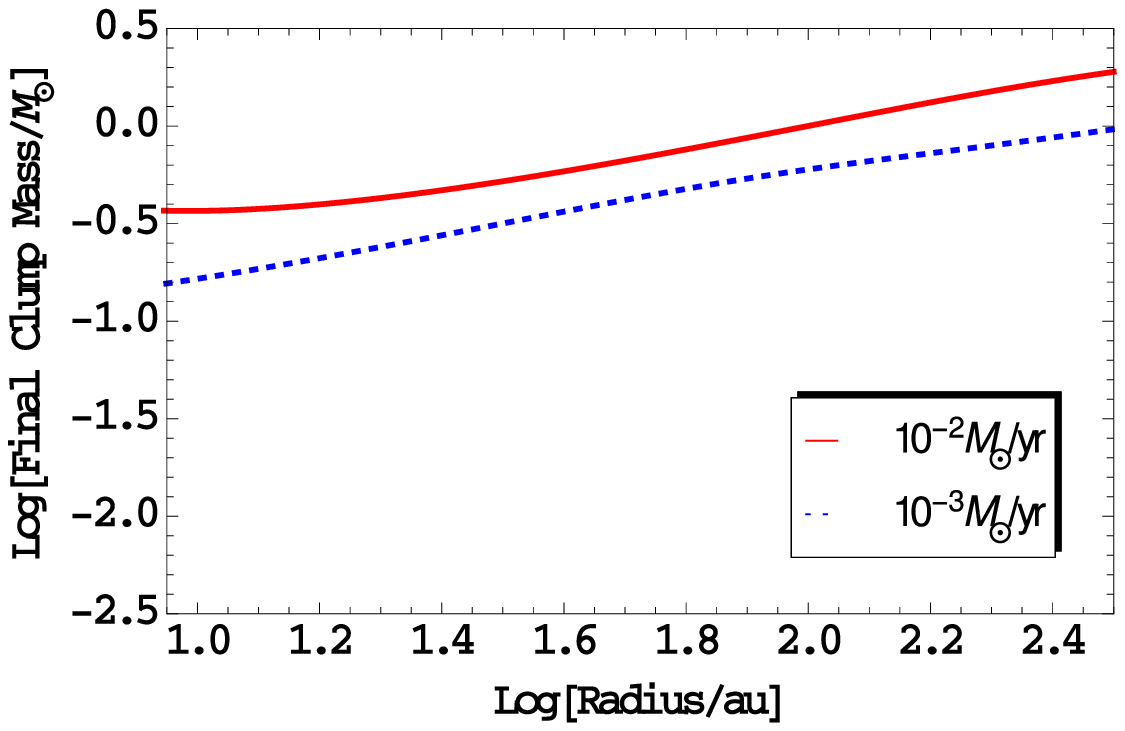}
\end{minipage}
\end{tabular}
\caption{The properties of the disk are plotted against the radius for the disk mass dominated case. The red and blue lines represent accretion rates of $\rm 10^{-2}~M_{\odot}$/yr and $\rm 10^{-3}~M_{\odot}$/yr respectively. The left panel shows the CIE cooling regime while the right panel depicts $\rm H_2$ cooling regime. The vertical magenta line shows the fragmentation radius and the horizontal green line marks the point where $\alpha$ = 1. The viscous parameter $\alpha$, the initial clump masses, accretion rates onto the clumps and their masses after the accretion are shown here.}
\label{fig1}
\end{figure*}

\begin{figure*}
\hspace{-6.0cm}
\centering
\begin{tabular}{c c}
\begin{minipage}{6cm}
\vspace{-0.1cm}
\includegraphics[scale=0.7]{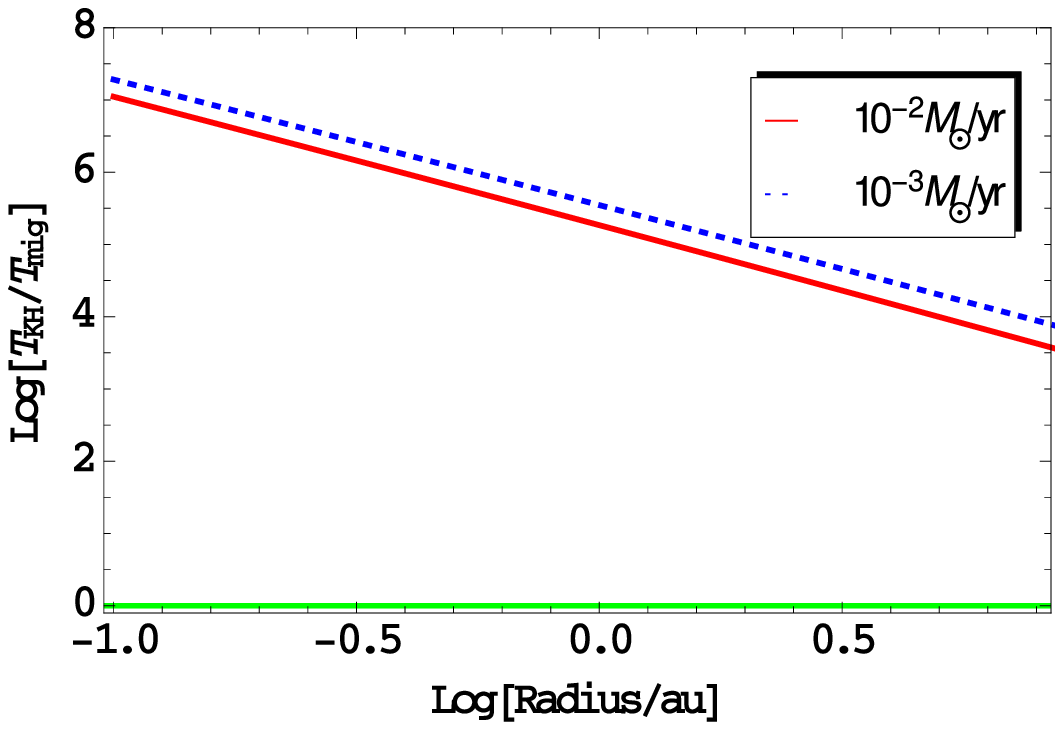}
\end{minipage}&
\begin{minipage}{6cm}
 \hspace{2cm}
\includegraphics[scale=0.7]{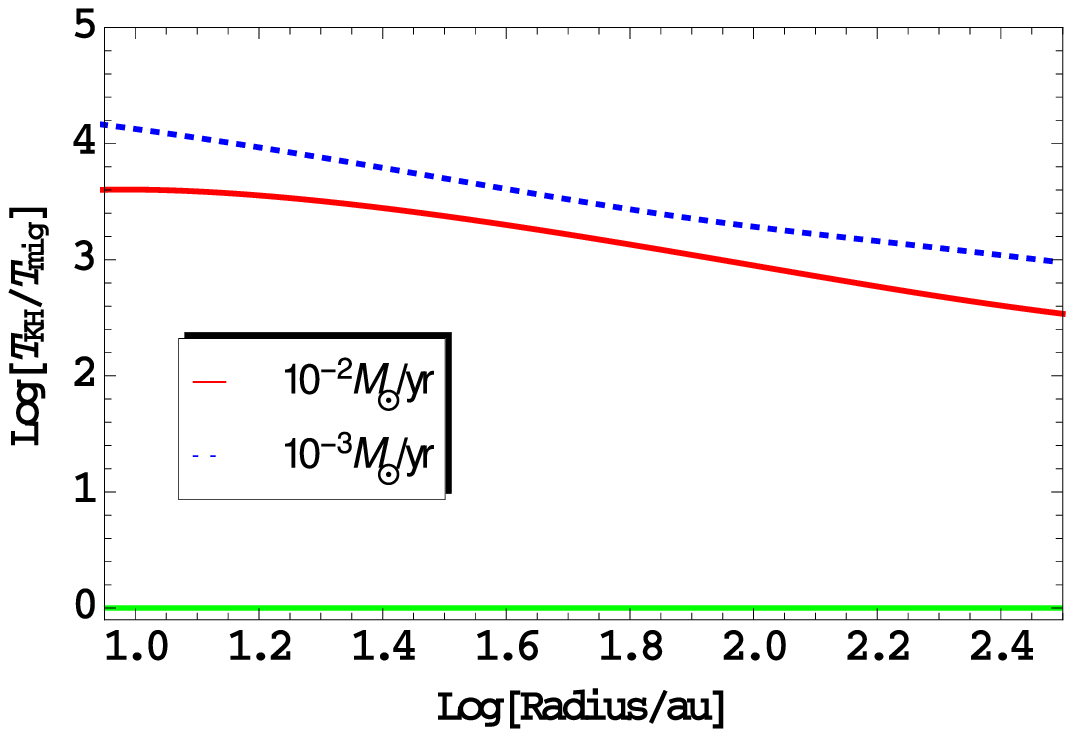}
\end{minipage} 
\end{tabular}
\caption{The ratio of the Kelvin Helmholtz to the migration time scales is shown here for the disk mass dominated case. The red and blue lines represent accretion rates of $\rm 10^{-2}~M_{\odot}$/yr and $\rm 10^{-3}~M_{\odot}$/yr respectively. The left panel shows the CIE cooling regime while the right panel depicts $\rm H_2$ cooling regime.}
\label{fig2}
\end{figure*}

\section{Results}
We present our results in this section. In the following subsections, we describe the properties at the initial stage of the collapse where most of the mass lies inside the disk. We also consider the late stage of disk evolution when central star is formed and investigate how it changes the disk structure. The potential impact of the stellar UV feedback on the disk is explored in the last section.

\subsection{The initial stage of fragmentation (disk-dominated)}
The properties of a marginally stable disk are computed by solving the set of equations presented in the previous section and are shown in figure \ref{fig}. We consider here typical accretion rates of $\rm 10^{-2}~M_{\odot}/yr$ (hereafter called $\rm \dot{M}_{-2}$) \& $\rm 10^{-3}~M_{\odot}/yr$ (hereafter called $\rm \dot{M}_{-3}$) often found in numerical simulations. For cooling, we consider optically thick $\rm H_2$ line cooling between densities of $\rm 10^{9}-10^{14}~cm^{-3}$ and above $\rm 10^{14}~cm^{-3}$ cooling due to the CIE. These are the major cooling mechanisms in in these regimes. The thermal evolution of the disk is determined by solving the energy conservation equation($\rm Q_+ = Q_-$) and is shown against the disk radius. As the density inside the disk increases, $\rm H_2$ line cooling becomes optically thick and the gas is heated up to 3100 K around 10 AU while the temperature declines towards larger radii. The temperature of the disk at 300 AU is about 400 K. It is found that the temperature of the disk is almost two times higher for $\rm \dot{M}_{-2}$ compared to $\rm \dot{M}_{-3}$. Towards smaller radii, the optically thick $\rm H_2$ cooling becomes inefficient, and therefore other cooling channels need to be considered.

In particular, the CIE cooling becomes effective and reduces the disk temperature to about 1400 K for $\rm \dot{M}_{-2}$ and 1000 K for $\rm \dot{M}_{-3}$. The particle density in the disk varies from $\rm 10^{10}-10^{14}~cm^{-3}$ between 10-300 AU and $\rm 10^{14}-10^{17}~cm^{-3}$ within the central 10 AU. The density profile follows $R^{-1.5}$ in the inner 10 AU and $R^{-2.6}$ between 10-300 AU. The density is about a factor of two higher for $\rm \dot{M}_{-2}$. The surface density is shallower within 0.1-10 AU ($\rm 10^4-5 \times 10^5 ~g/cm^{2}$) and becomes steeper between 10-300 AU as it varies from $\rm 30-10^4 ~g/cm^{2}$. The mass of the disk at a given radius is shown in the bottom panel of figure \ref{fig}. The disk is about two times more massive for $\rm \dot{M}_{-2}$ in comparison with  $\rm \dot{M}_{-3}$. 

Motivated by the hydrodynamical simulations and analytical studies which show that the disk becomes unstable for $ \alpha > \alpha_{\rm crit}$ \citep{Gammie01,Rice05,Inayoshi2014}, \cite{Zhu2012} investigated the fragmentation in a protoplanetary disk by performing numerical simulations and show that the disk becomes unstable for $\alpha > 1.0$. To measure the stability of the disk we use $\alpha > 1.0$ in our model as a fragmentation criterion but also discuss below how this choice affects our results. We hereafter call this point the fragmentation radius. We have computed the value of $\alpha$ for the disk and it is shown in figure \ref{fig1}. The results from our model show that the value of $\alpha$ remains lower than 1 inside the central 10 AU for both accretion rates. At larger radii, $\alpha$ remains lower than one for $\rm \dot{M}_{-3}$ but becomes larger than one for $\rm \dot{M}_{-2}$ around 100 AU.  Based on these results, it is expected that the disk remains stable for smaller accretion rates (i.e. $\rm \dot{M}_{-3}$) and becomes unstable at larger radii for  $\rm \dot{M}_{-2}$. However, for $0.06 < \alpha_{crit} <1.0$ (see our discussion below) the disk becomes unstable and may even fragment at smaller radii. In fact, high resolution simulations show fragmentation on scales of less than 1 AU but it is not clear whether these clump will form stars as the simulations were evolved only for very short time scales \citep{Greif12}.

We have also estimated the  masses of the clumps which are shown in figure \ref{fig1}. They have sub-solar masses and are more massive in the outer parts of the disk than inside. The typical accretion rates on the clumps shown in figure \ref{fig1} are about $\rm 10^{-3}~M_{\odot}/yr$ and about an order of magnitude higher for $\rm \dot{M}_{-2}$. The clumps grow via accretion from the ambient medium and their masses are increased by an order of magnitude. The final masses of the clumps are shown in the bottom panel of figure \ref{fig1}. It is found that final clump masses in the outskirts of the disk reach about a solar mass for $\rm \dot{M}_{-2}$. To know the fate of these clumps, we have computed their migration time and compared it with the Kelvin-Helmholtz contraction time scale ($\rm T_{KH}$). The comparison is shown in figure \ref{fig2}, it is found that the migration time is shorter than the $\rm T_{KH}$. Therefore, clumps will migrate inward before they reach the main sequence and produce UV feedback. 

\begin{figure*}
\hspace{-6.0cm}
\centering
\begin{tabular}{c c}
\begin{minipage}{6cm}
\vspace{-0.6cm}
\includegraphics[scale=0.7]{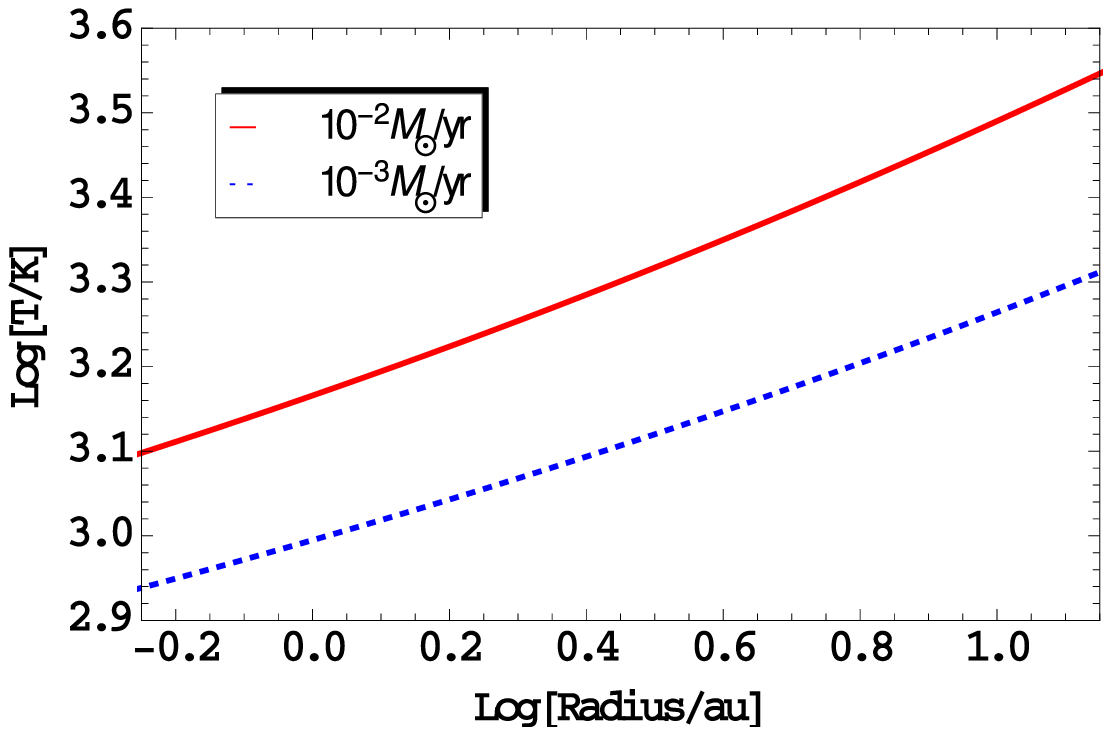}
\end{minipage}&
\begin{minipage}{6cm}
\hspace{2.53cm}
\includegraphics[scale=0.7]{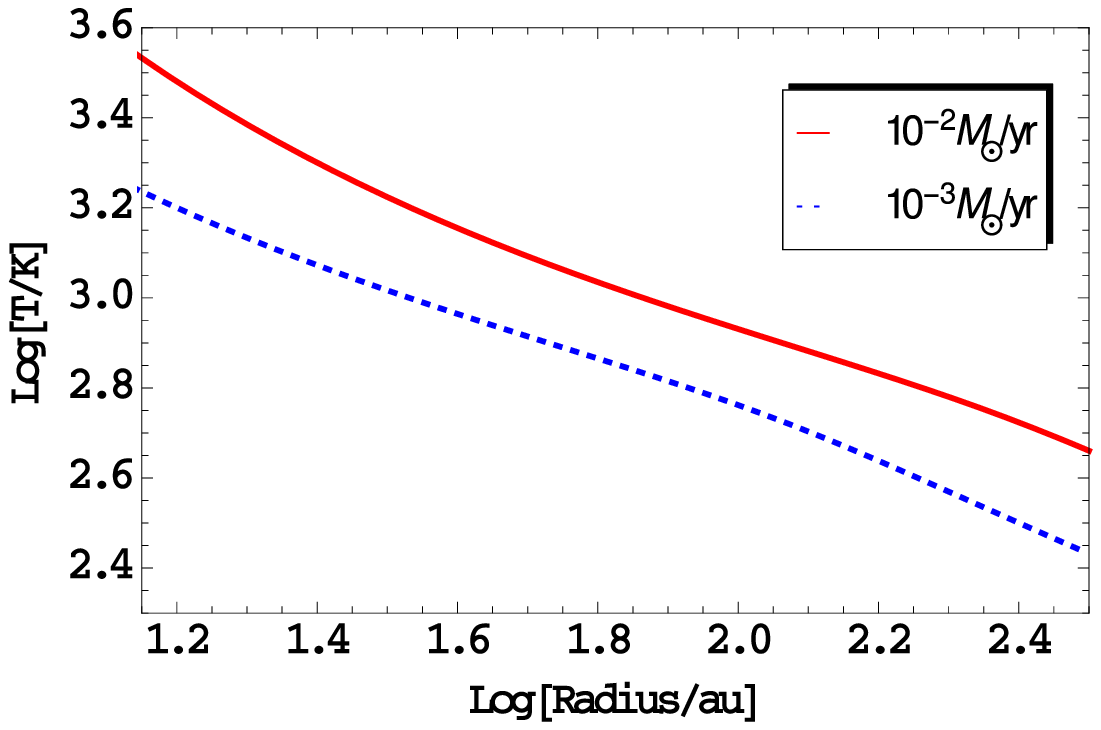}
\end{minipage} \\
\begin{minipage}{6cm}
\vspace{-0.6cm}
\includegraphics[scale=0.7]{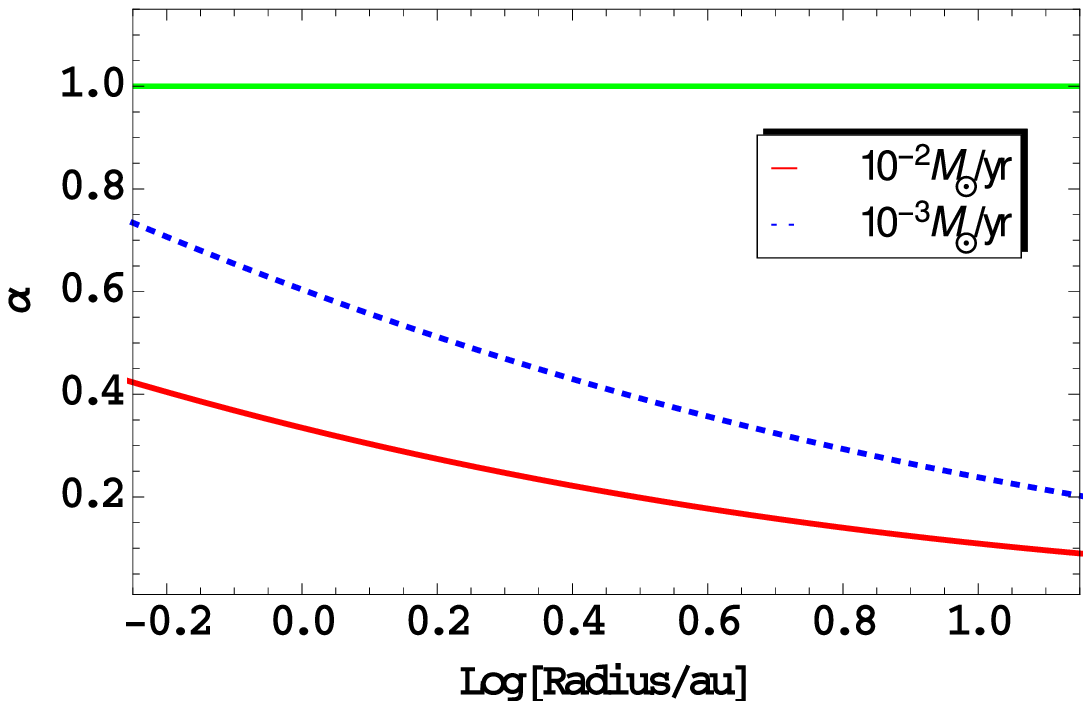}
\end{minipage}&
\begin{minipage}{6cm}
\hspace{2.6cm}
\includegraphics[scale=0.7]{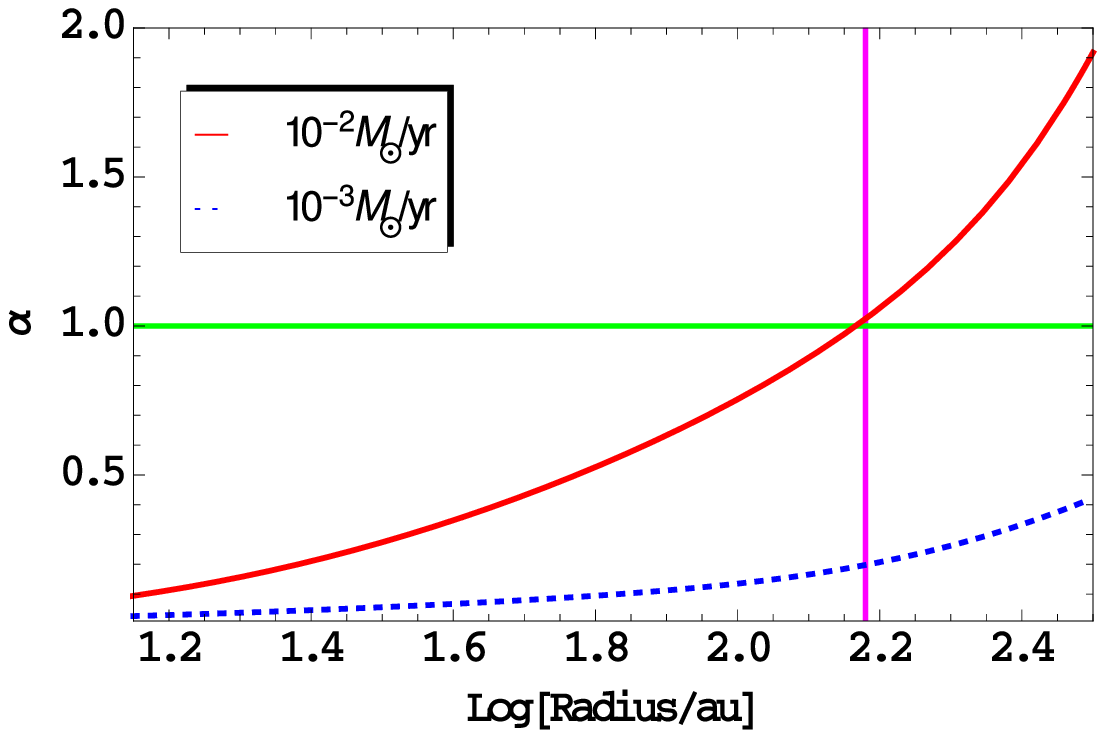}
\end{minipage} \\
\begin{minipage}{6cm}
\vspace{-0.6cm}
\includegraphics[scale=0.7]{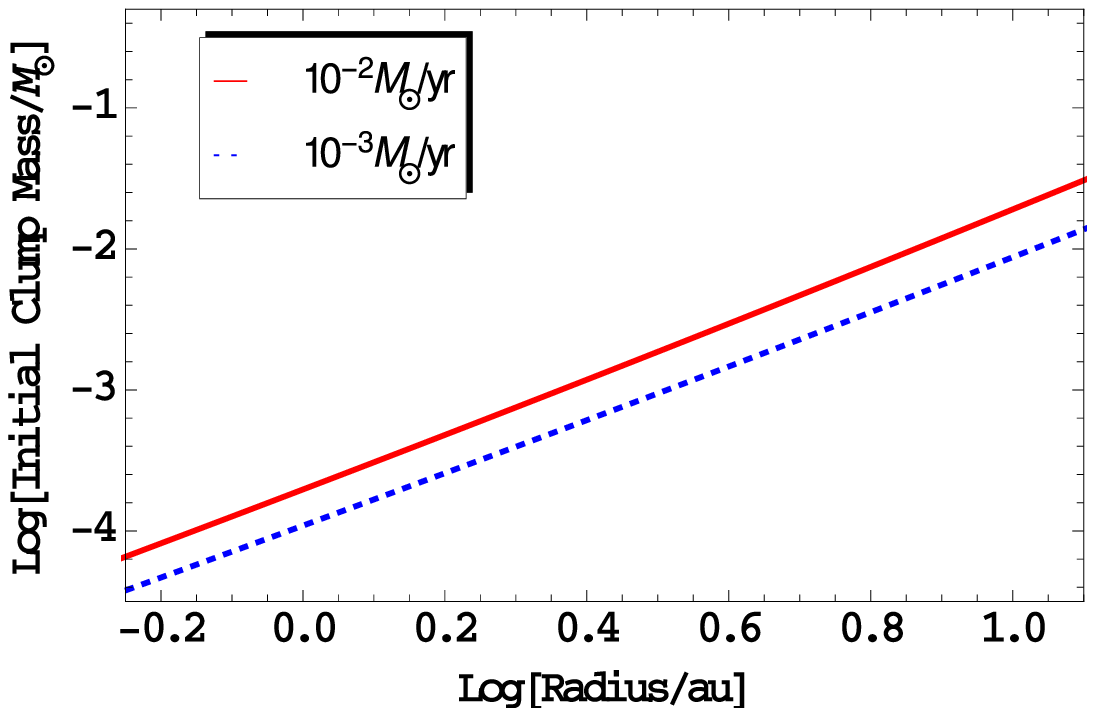}
\end{minipage}&
\begin{minipage}{6cm}
 \hspace{2cm}
\includegraphics[scale=0.7]{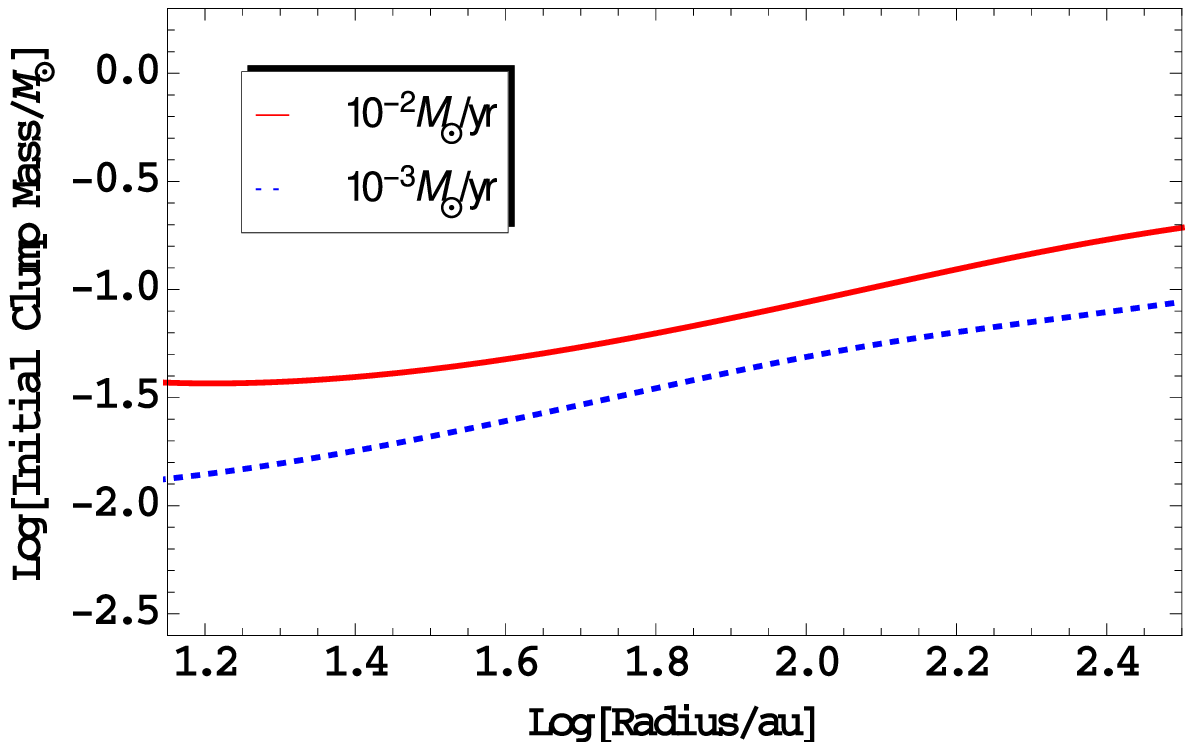}
\end{minipage} \\
\begin{minipage}{6cm}
\vspace{-0.6cm}
\includegraphics[scale=0.7]{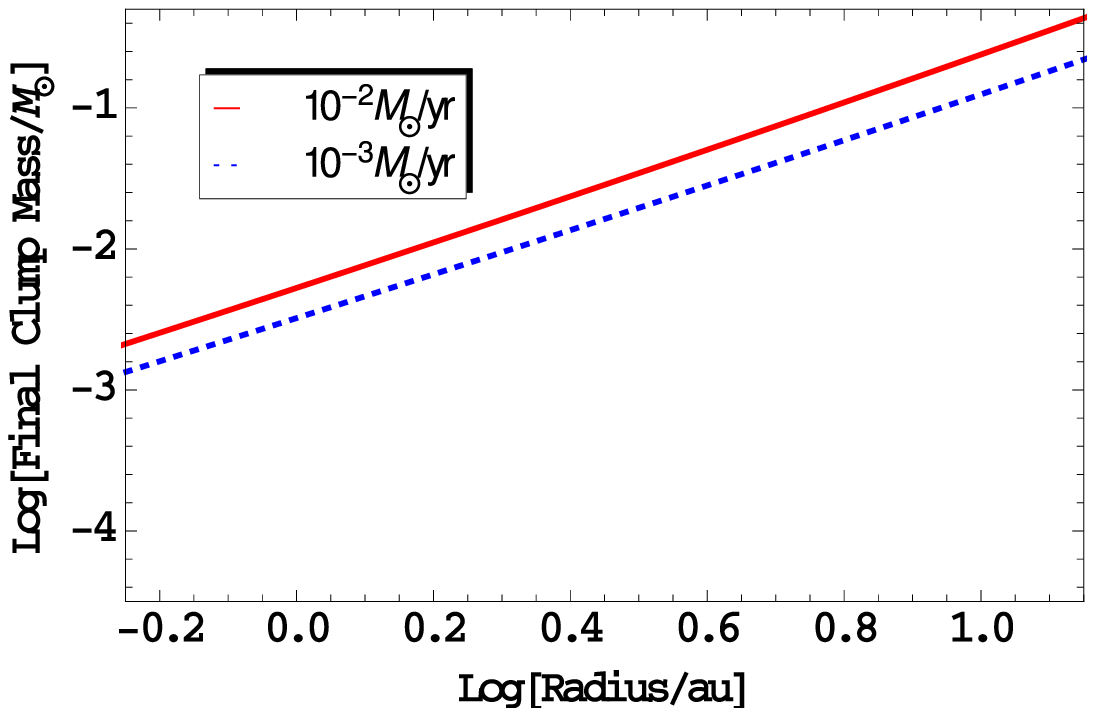}
\end{minipage}&
\begin{minipage}{6cm}
 \hspace{2.4cm}
\includegraphics[scale=0.7]{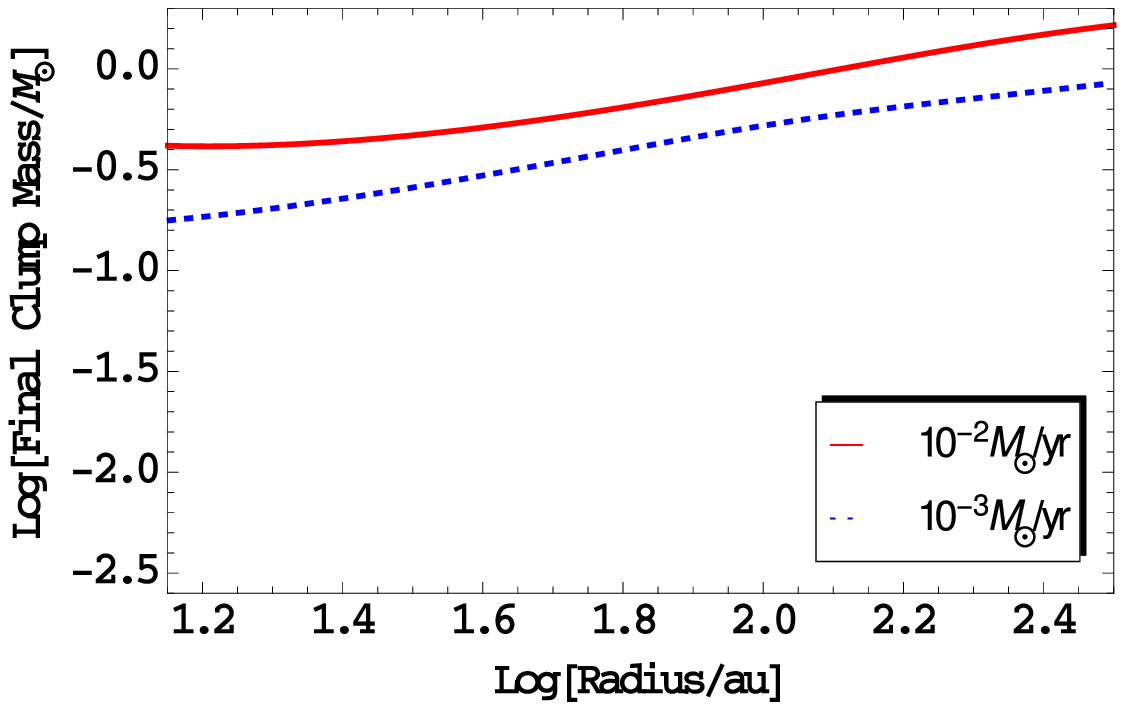}
\end{minipage} 
\end{tabular}
\caption{The properties of the disk are plotted against the radius for the central star of 5 solar masses. The red and blue lines represent accretion rates of $\rm 10^{-2}~M_{\odot}$/yr and $\rm 10^{-3}~M_{\odot}$/yr respectively. The vertical magenta line shows the fragmentation radius and the horizontal green line marks the point where $\alpha$ = 1. The left panel shows the CIE cooling regime while the right panel depicts $\rm H_2$ cooling regime. The temperature, the viscous parameter $\alpha$, the clump masses before and after the accretion are shown here.}
\label{fig4}
\end{figure*}

\begin{figure*}
\hspace{-6.0cm}
\centering
\begin{tabular}{c c}
\begin{minipage}{6cm}
\vspace{-0.1cm}
\includegraphics[scale=0.7]{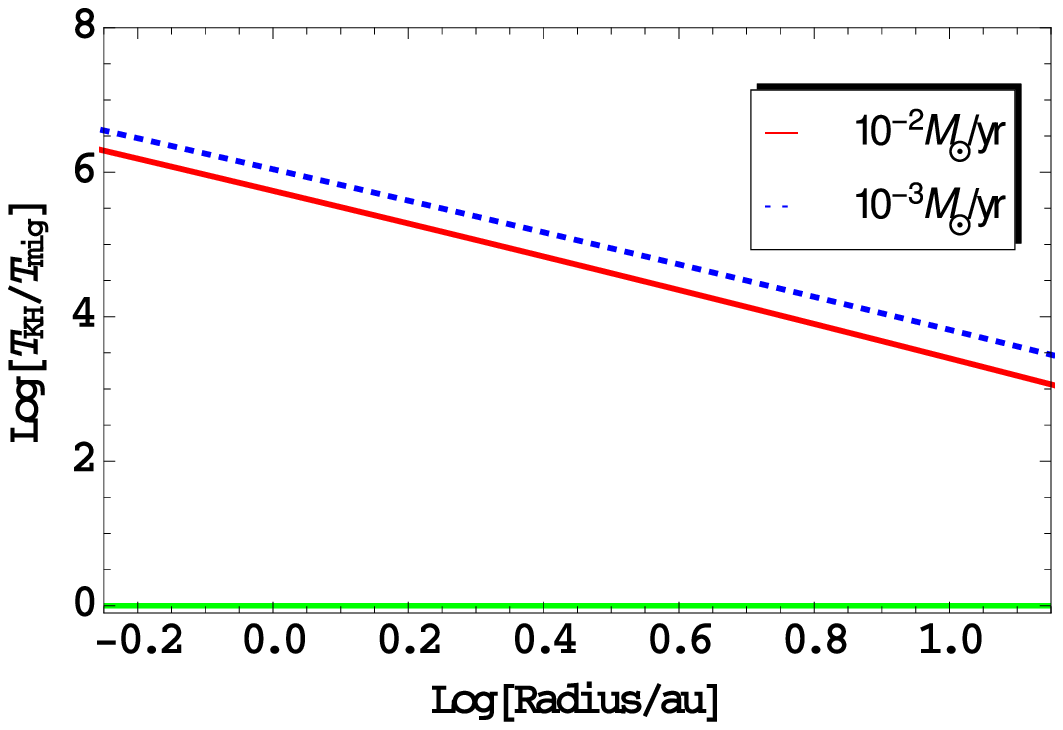}
\end{minipage}&
\begin{minipage}{6cm}
\hspace{1.8cm}
\includegraphics[scale=0.7]{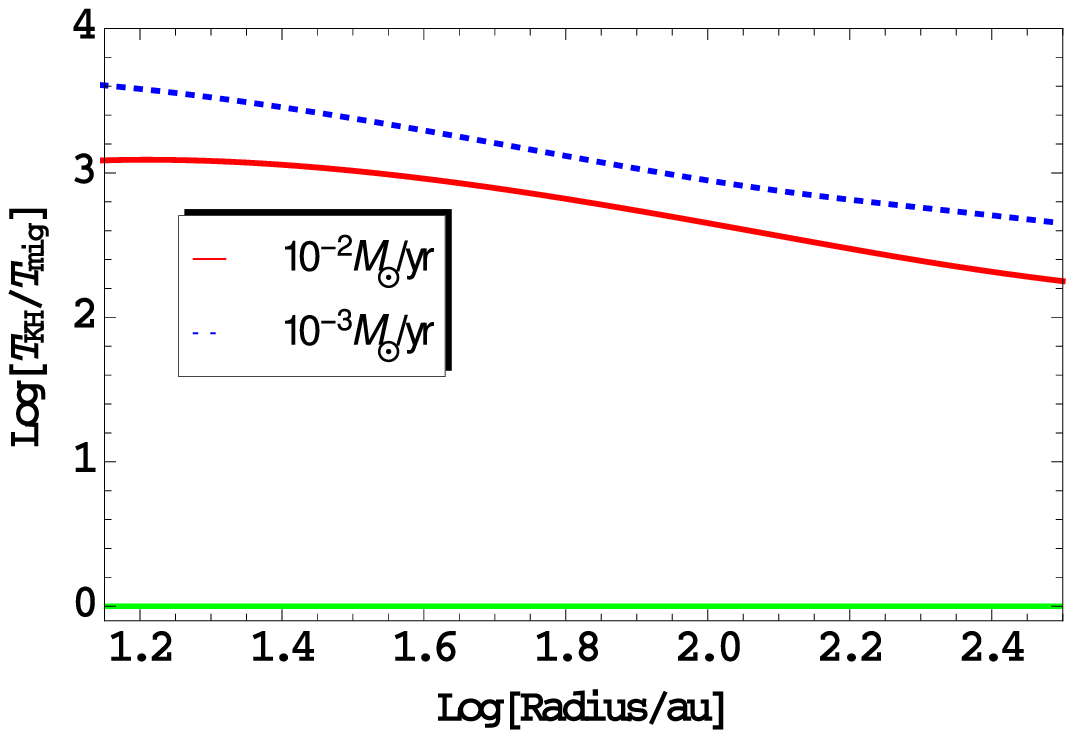}
\end{minipage}
\end{tabular}
\caption{The ratio of the Kelvin Helmholtz to the migration time scales is shown for the central star of 5 solar masses. The red and blue lines represent accretion rates of $\rm 10^{-2}~M_{\odot}$/yr and $\rm 10^{-3}~M_{\odot}$/yr respectively. The left panel shows the CIE cooling regime while the right panel depicts $\rm H_2$ cooling regime.}
\label{fig41}
\end{figure*}

\begin{figure*}
\hspace{-6.0cm}
\centering
\begin{tabular}{c c}
\begin{minipage}{6cm}
\vspace{-0.6cm}
\includegraphics[scale=0.7]{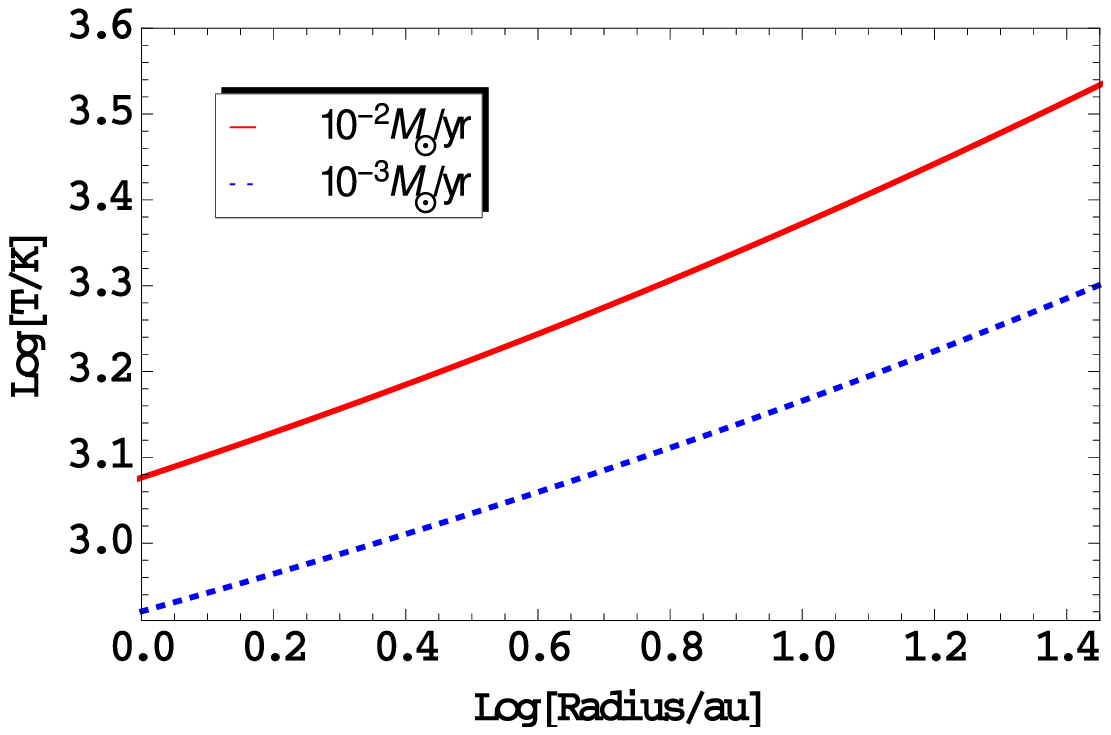}
\end{minipage}&
\begin{minipage}{6cm}
 \hspace{2cm}
\includegraphics[scale=0.7]{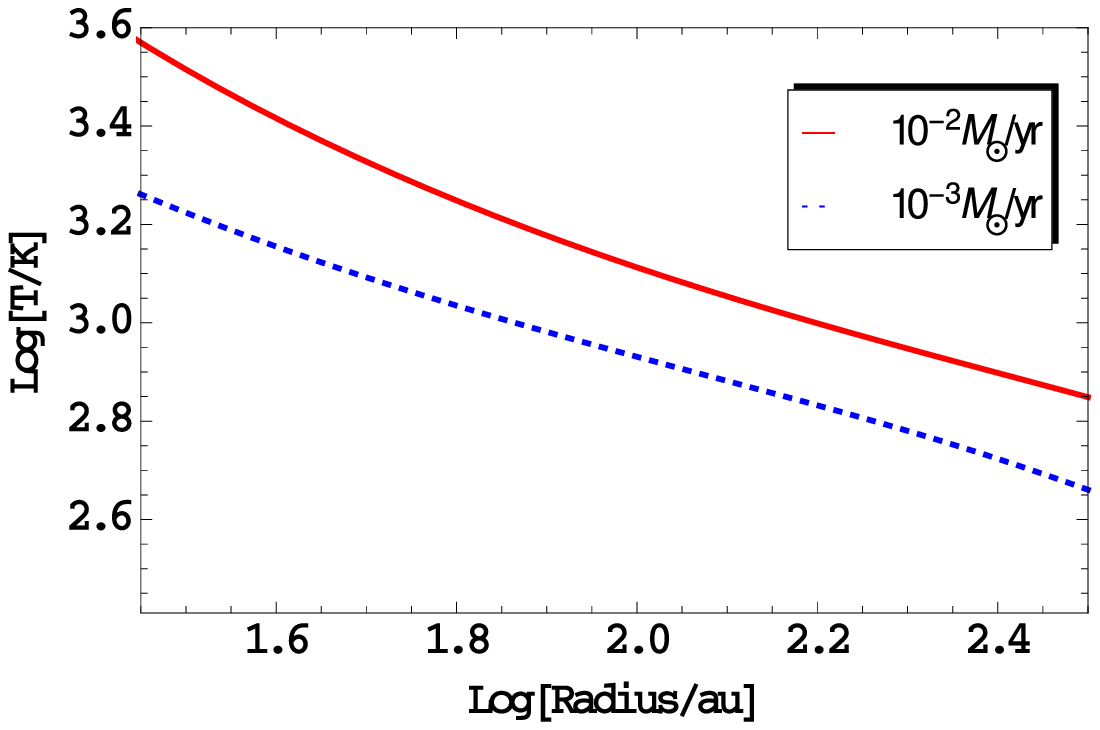}
\end{minipage} \\
\begin{minipage}{6cm}
\vspace{-0.6cm}
\includegraphics[scale=0.7]{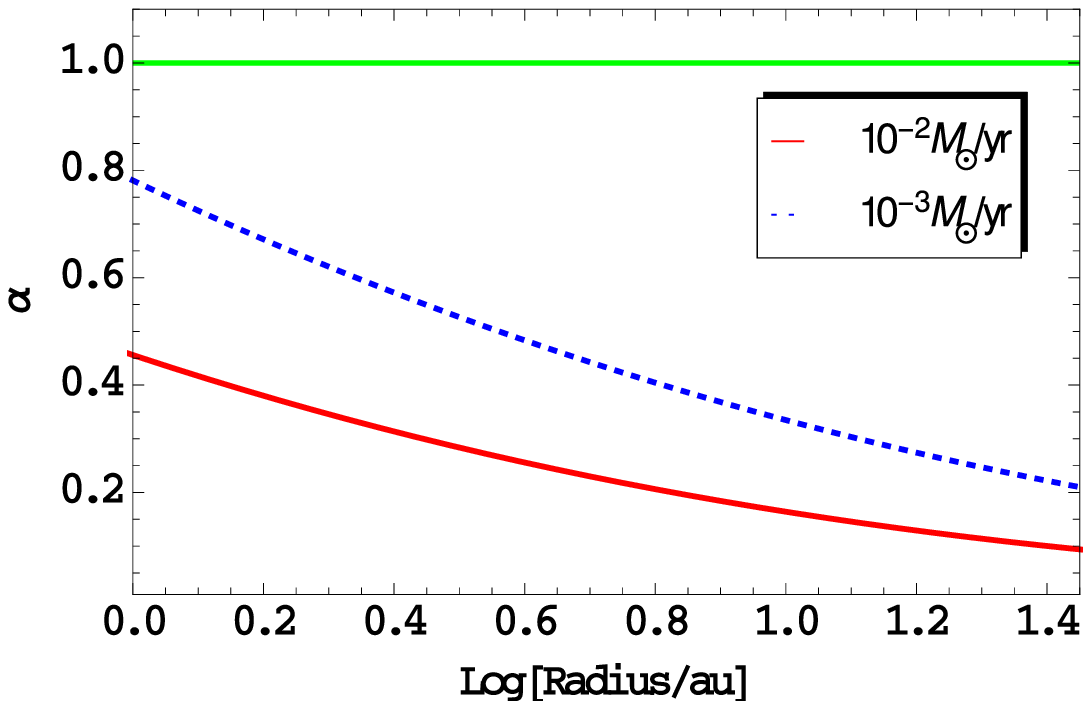}
\end{minipage}&
\begin{minipage}{6cm}
 \hspace{2cm}
\includegraphics[scale=0.7]{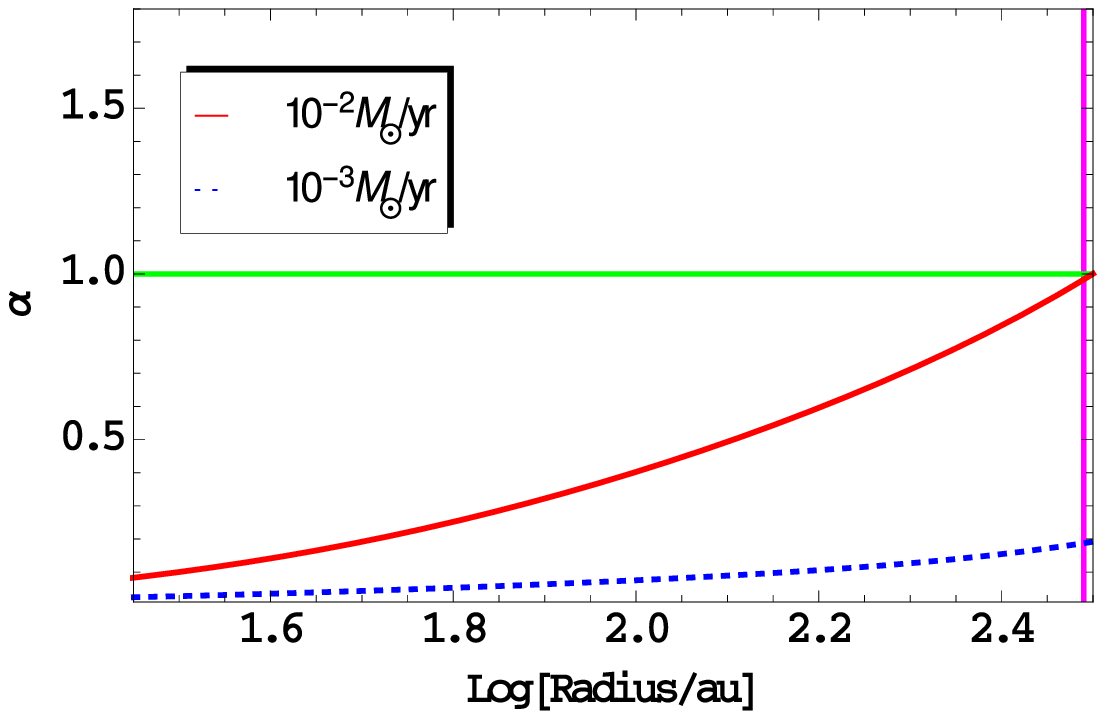}
\end{minipage} 
\end{tabular}
\caption{The temperature of the disk and the viscous parameter $\alpha$ are plotted against the radius for the central star of 50 solar masses. The red and blue lines represent accretion rates of $\rm 10^{-2}~M_{\odot}$/yr and $\rm 10^{-3}~M_{\odot}$/yr respectively. The vertical magenta line shows the fragmentation radius and the horizontal green line marks the point where $\alpha$=1. The left panel shows the CIE cooling regime while the right panel depicts $\rm H_2$ cooling regime.}
\label{fig5}
\end{figure*}

\begin{figure*}
\hspace{-6.0cm}
\centering
\begin{tabular}{c c}
\begin{minipage}{6cm}
\vspace{-0.6cm}
\includegraphics[scale=0.7]{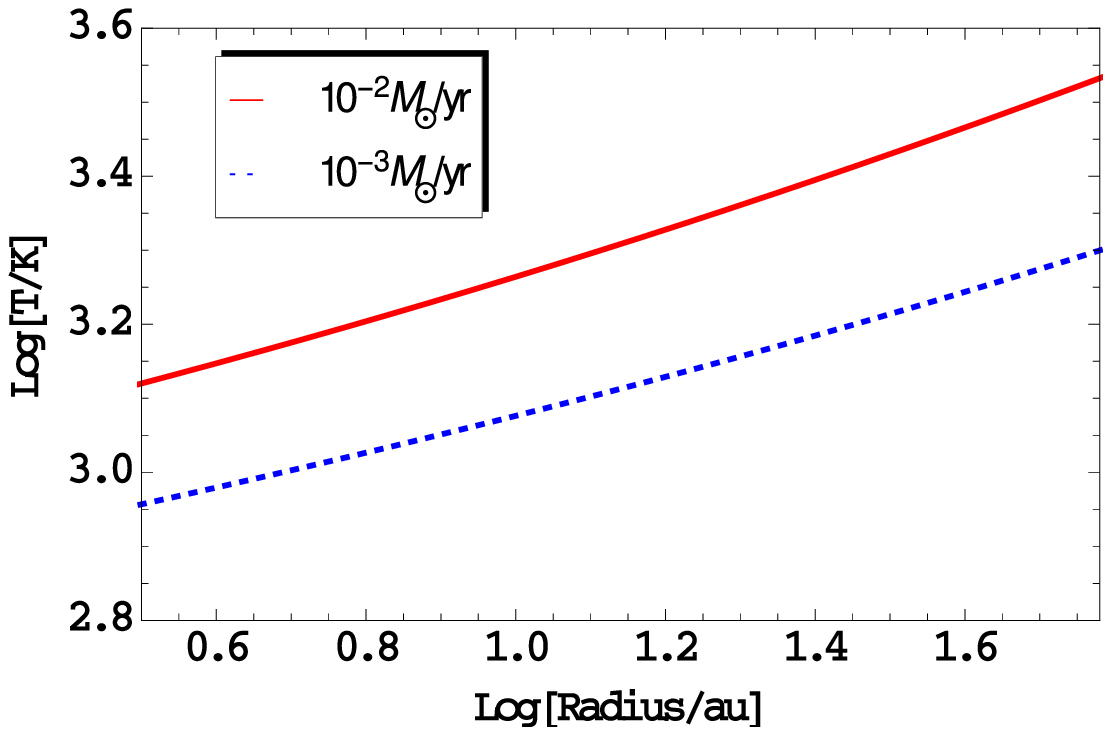}
\end{minipage}&
\begin{minipage}{6cm}
 \hspace{2cm}
\includegraphics[scale=0.7]{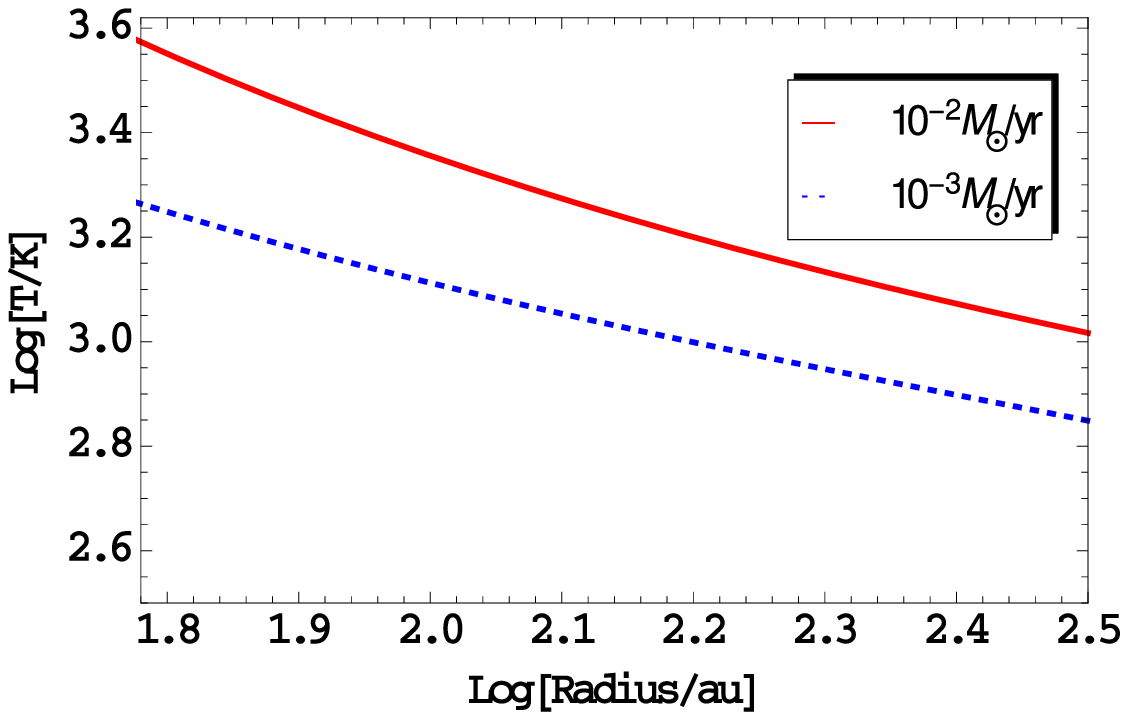}
\end{minipage} \\
\begin{minipage}{6cm}
\vspace{-0.6cm}
\includegraphics[scale=0.7]{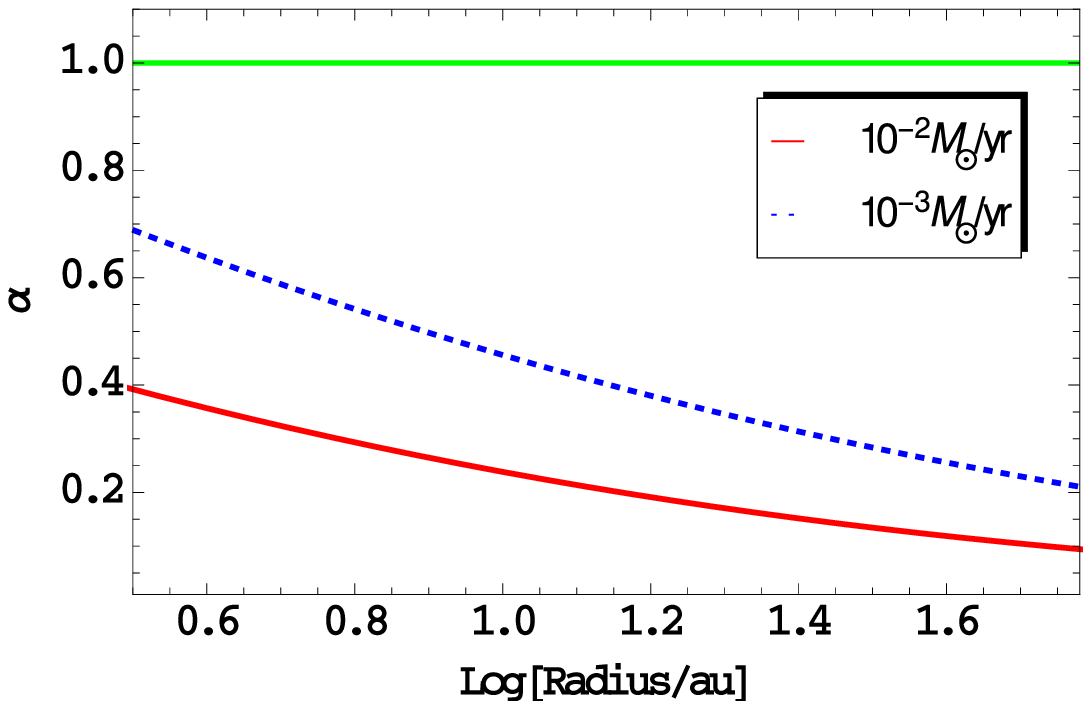}
\end{minipage}&
\begin{minipage}{6cm}
 \hspace{2cm}
\includegraphics[scale=0.7]{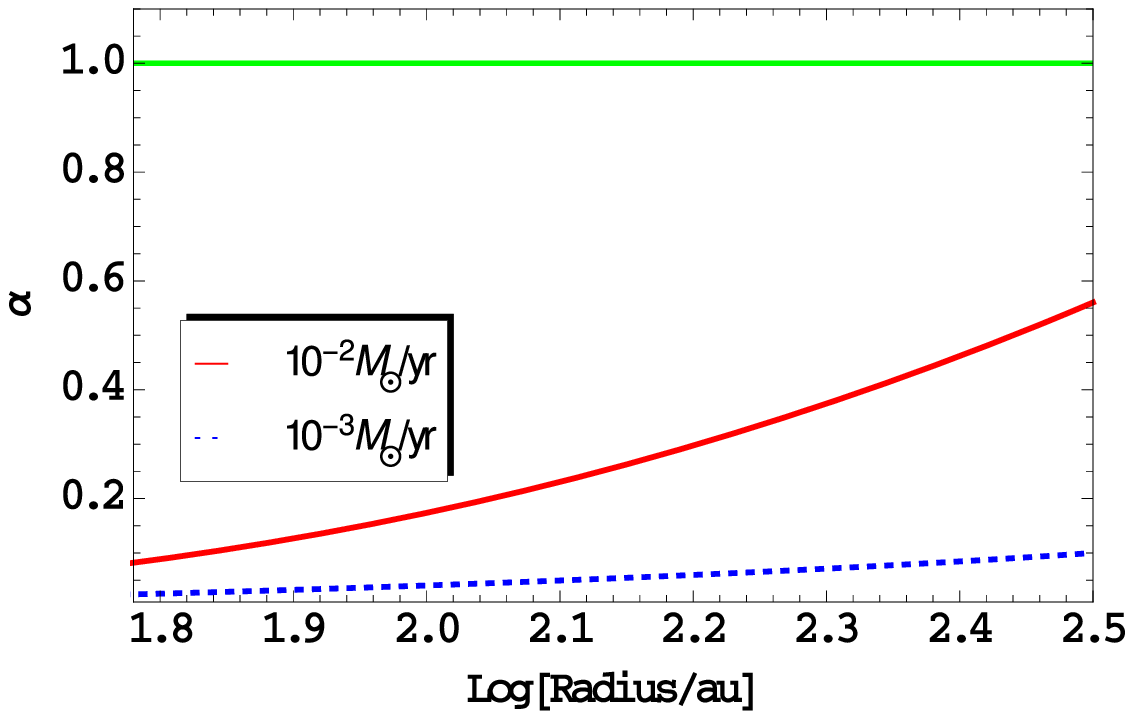}
\end{minipage} 
\end{tabular}
\caption{The temperature of the disk  and the viscous parameter $\alpha$ are plotted against the radius for the central star of 500 solar masses. The red and blue lines represent accretion rates of $\rm 10^{-2}~M_{\odot}$/yr and $\rm 10^{-3}~M_{\odot}$/yr respectively. The horizontal green line shows the point where $\alpha$=1. The left panel shows the CIE cooling regime while the right panel depicts $\rm H_2$ cooling regime.}
\label{fig6}
\end{figure*}


\subsection{Fragmentation in the presence of a central object (star-dominated)}
In the later stages of the disk evolution, it is expected that the potential from the central star starts to dominate and changes the disk structure. We have estimated the disk properties for a central star of 5, 50 and 500 solar masses. For the accretion rates considered here, the expected radii of these star are smaller than 100 $\rm R_{\odot}$ \citep{Hosokawa12,Schleicher13}. Therefore, we do not show the scales smaller than this in our model for the star dominated cases. It is found that in the presence of the central star, the disk becomes denser and hotter. Therefore, the CIE cooling kicks in earlier at larger radii as shown in figure \ref{fig4}. The density of $\rm 10^{17}~cm^{-3}$ is already reached at 0.6 AU where the gas start to become  optically thick to CIE cooling which is not included in our model. However, we expect that at higher densities the temperature in the disk begins to evolve almost adiabatically. The value of the viscous parameter $\alpha$ remains lower than 1 for $\rm \dot{M}_{-3}$ similar to the disk dominated case. For $\rm \dot{M}_{-2}$, the disk becomes unstable above 100 AU and remains stable in the interior.

The masses of the clumps are similar to the disk dominated cases both in the $\rm H_2$ line and CIE cooling regimes. About solar mass clumps are expected to form between 100-300 AU where the disk becomes unstable. We have also compared the migration and Kelvin-Helmholtz time scales as shown in figure \ref{fig41}. It is found that the ratio of $\rm T_{mig}$ to $\rm T_{KH}$ remains close to one. It comes from the fact that in the star mass dominated case, the orbital period becomes shorter and the ratio of  $\rm T_{KH}$ to $\rm T_{mig}$ declines.

We have also considered the cases with a central star of 50 and 500 solar masses. The disk becomes denser and warmer even at larger radii compared to the disk mass dominated case as shown in figures \ref{fig5} \& \ref{fig6}. For the case of 50 solar mass central star and $\rm \dot{M}_{-2}$, the fragmentation radius  shifts further to larger radii of 300 AU. The disk becomes completely stable for both accretion rates for the central star of 500 $\rm M_{\odot}$. Moreover, the migration time scale remains shorter than KH time scale. Therefore, the clumps are expected to migrate inward. Our estimates for the Roche limit for the central star of 500 solar masses show that clumps will not be able to survive within the 1 AU due to the tidal forces exerted by the central star.

\subsection{Dependence on the critical value of alpha}
So far we have assumed the value of $\alpha_{crit}$ = 1 as the criterion for the stability of the disk. This choice is motivated by the works of \cite{Zhu2012} \& \cite{Inayoshi2014}. However, the critical value of $\alpha$ can vary between 0.06 and 1. \cite{Rice05} have suggested that the value of $\alpha_{crit}$ is 0.06 although they did not have the realistic model for the heating and cooling of the disk. By considering the critical value of $\alpha$ = 0.06, the disk in our model becomes unstable for both accretion rates and will fragment. Fig. 2 in \cite{Zhu2012} shows that in fact realistic cases may lie in between these extreme scenarios, so it is possible that fragmentation occurs for the value of $\alpha$ smaller than 1. Therefore, fragmentation may occur in the disk interior as found in high resolution numerical simulations \citep{Greif12}. Nevertheless, the ratio of $\rm T_{KH}$ to $\rm T_{mig}$ always remains $\rm \geq 1$. The clumps are expected to migrate inward before reaching the main sequence. Furthermore, they will be tidally disrupted in the central part of the disk depending on the mass of the central star. Therefore, we argue that the choice of the critical value of $\alpha$ does not change our main conclusions.

\subsection{Radiative feedback from the central object}
Once the central star reaches the main sequence, it will start to produce UV flux which can photo-ionize the gas in the surroundings and photo-evaporate the disk. The studies which consider the impact of stellar UV feedback onto the accretion flows through disk show that effect of UV starts to become important about 20 $\rm M_{\odot}$ and photo-evaporation shuts the accretion between 50-100 $\rm M_{\odot}$ \citep{Mckee2008,Hosokawa11}. However, the accretion from the disk may still continue. To compute the photo-evaporation rate of the disk, we use the expression given in \cite{Hollenbach94}:
\begin{equation}
\dot{M}_{PE} = 1.3 \times 10^{-5} \left(\frac{\phi_{UV}}{10^{49}~s^{-1}} \right)^{1/2}  \left(\frac{M_{*}}{10~M_{\odot}}\right)^{1/2}~M_{\odot}/yr 
\label{eq22}
\end{equation} 
here $\phi_{{\rm UV}}$ is the photo-ionizing rate and $\rm M_{*}$ is the mass of a star. A star of 40 $\rm M_{\odot}$ produces $\phi_{{\rm UV}}$$\rm = 1.8 \times 10^{49}~s^{-1}$ \citep{Schaerer2002}. It gives the photo-evaporation rate of $\rm 3.5 ~ \times ~10^{-5}~M_{\odot}/yr$. For a 500 $\rm M_{\odot}$, $\phi_{\rm UV}$$\rm =5.2 \times 10^{50}~s^{-1}$ which gives photo-evaporation rate of $\rm 2.0 ~ \times ~10^{-4}~M_{\odot}/yr$. These are about an order of magnitude lower than the adopted mass accretion rates. Therefore, we expect that photo-evaporation will not be able to completely terminate the accretion provided that such large accretion rates of the order of $\rm \dot{M}_{-2}$ are maintained. However, \cite{Hosokawa11} show that accretion rate declines down when star reaches above 40 solar masses. We expect that the latter is partly due to hydrodynamical effects and angular momentum conservation (see also \cite{2013MNRAS.436.2989L}), but also the UV feedback in the vertical direction, which decreases the mass supply onto the disk. 

Recently \cite{Tanak2013} computed the photo-evaporation rate from the disk by performing 2-D asymmetric radiative transfer calculations and found that the photo-evaporation rate depends on the disk radius instead of the gravitational radius. We also checked for the expression given in their equation 24 and found similar results. 

To further assess the impact of photo-ionization, we compute the size of HII region by assuming the equilibrium between ionization and recombination. The equilibrium condition is given as: 
\begin{equation}
\dot{N} = \int_{R_{i}}^{R_S} 4 \pi \alpha_{rec}  r^{2} (n_{0}(r/R)^{l})^{2} dR
\label{eq23}
\end{equation}
here N is the photo-ionization rate, $\rm R_S$ is the radius of HII region, $\rm R_i$ is inner radius of the disk, $\rm n_{0}$ is the corresponding density, $\alpha_{rec}$ is the recombination rate for hydrogen and $\ell$ is the exponent of the density profile. Solving the above equation for $\rm R_S$, we get the following expression:

\begin{equation}
R_S = \left({R_{i}}^{3-2l} + (3-2l) \frac{\dot{N}}{4 \pi \alpha_{rec} n_{0}^2R^{2l} } \right)^{1/(3-2l)}
\label{eq24}
\end{equation}
Taking the numbers found in our model, $\rm n_0 = 10^{17}~cm^3$, $\alpha_{rec} = {\rm 3 .0 \times 10^{-13}~cm^3/s}$, $\rm R = 0.1~AU$ and ${\ell \rm = 1.6}$ and R$_{\rm i}$ $\rm = 10~R_{\odot}$. For a 40 solar mass star ($\rm \dot{N}~=~ 10^{49}~s^{-1}$) the size of HII region is 0.04 AU. It is smaller than the scales probed in our model. We further expect that ionizing feedback from the central star of about 100 $\rm M_{\odot}$ will not be able to strongly influence the disk structure. This result is further consistent with \cite{Hosokawa11} who also found for a star $\rm > 40~M_{\odot}$ the HII region develops in the vertical direction and later on start to expand horizontally.

Our approach does not include the geometrical effects and assumes constant accretion rates. As found from the simulations, the declining accretion rates and the photo-evaporation of the disk is expected to shut down the accretion above 100 solar masses.

\section{Discussion}

We present here a simple analytical model to study the fragmentation properties of a marginally stable steady-state disk. Our model assumes that the Toomre Q=1 and the disk is fully molecular. We solve the set of equations to compute the viscous parameter $\alpha$. We check the stability of the disk by employing the fragmentation criterion based on recent numerical simulations which show that the disk becomes unstable for $\alpha {\rm ~ > 1}$. Our model includes a recipe for optically thick $\rm H_2$ line cooling. We further take into account the CIE cooling occurring at high densities. The disk temperature is evaluated by assuming that heating by the turbulence viscosity and shocks is balanced by the cooling. The typical masses of the clumps are computed and their fate is determined by comparing the migration and contraction time scales.

Our findings show that the temperature inside the disk increases for higher accretion rates and below 10 AU the CIE cooling becomes effective which brings the temperature down to about 1000 AU. The densities in the disk range from $\rm 10^{10}-10^{17}~cm^{-3}$. Our model shows that the stability of the disk depends on the critical value of the $\alpha$. It remains stable for an accretion rate $\rm \leq 10^{-3}~M_{\odot}/yr$ and becomes unstable at radii $\rm > 100~AU$ in the presence of higher accretion rate of  $\rm  10^{-2}~M_{\odot}/yr$ for $\alpha > 1$. The typical masses of the clumps are about $\rm 0.1~M_{\odot}$ and reach about solar masses in the regime where disk is unstable. For $\alpha_{crit}$ between 0.06 and 1, the disk becomes unstable for both accretion rates. In this case, the disk is expected to fragment and form multiple clumps also on scales around 1 AU as seen in high resolution simulations \citep{Greif12}. To compute the fate of the clumps, we compared the migration and the Kelvin-Helmholtz contraction time scales. It is found that the clumps are migrated inward and merge with the central star before reaching the main sequence. The clumps will be tidally disrupted within the central 1 AU in the presence of a massive central star. So, the choice of the critical value $\alpha_{crit}$ is not going to affect our main conclusions


We have also considered a few cases of the later disk phase when matter is accreted into the center and a star is formed. In such a case, it dominates the central potential and changes the disk structure. We found that in the presence of a central star, the disk becomes denser and hotter. For a central star of 5 solar masses, the disk properties almost remain similar to the disk dominated case. However, for a star of 50 \& $\rm 500~M_{\odot}$ the fragmentation radius shifts outward and the disk becomes completely stable. Similar to the disk dominated case, clumps are expected to migrate inward and merge with central star. Our choice for the 50 \& 500 $\rm M_{\odot}$ stars comes from that fact that if accretion rates of $\rm 10^{-2}~M_{\odot}/yr$ are continued for 10,000 years (the typical time before reaching the main sequence) then a massive star of the above mentioned mass is expected to form. In fact, numerical simulations including the radiative feedback from the star show that such massive stars can form \citep{Hirano2014, Susa14}.

We assumed constant accretion rates  of $\rm 10^{-3}$ \& $\rm 10^{-2}~M_{\odot}/yr$ onto the disk and found that in the presence of such accretion rates, the photo-evaporation rate from the stellar UV feedback remains an order of magnitude lower. We therefore expect photo-evaporation to become relevant once the accretion rate drops below $\rm 2 \times 10^{-4} M_{\odot}/yr$. Similarly, we found that size of HII region remains smaller than 1 AU and would not be able to completely disrupt the disk in equatorial plane. So, we expect that a star of about 100 $\rm M_{\odot}$ may form. However, a HII region is likely to develop in the polar direction which will further expand with time as shown in numerical simulations \citep{Hosokawa11, Stacy2012}. This may result in a decline of the accretion rates with time. \cite{Mckee2008} have estimated the impact of UV feedback from the star by including the radiation pressure and $\rm H_2$ photo-dissociation rate. They also found that  HII region develops in the polar direction for a 30 $\rm M_{\odot}$ star which further expands beyond the gravitational radius between 50-100 solar masses. However, the accretion from the equatorial plane may continue. According to their estimates, above 140 solar masses, the combination of declining accretion rates and photo-evaporation terminates the accretion. 

We have studied the case where the disk is of a primordial composition which is valid for the disks around Pop III stars. However, the disks forming around the second generation of stars may already be polluted by the metals and dust ejected by the supernova explosions. The presence of trace amount of dust and metals may change the disk structure and make it unstable. \cite{Tanaka2014} have studied the disk fragmentation in various environments from zero to solar metallicity by employing an analytical model. They used an alternative approach by assuming that the viscous parameter $\alpha$ = 0.1-1 and computed the Toomre Q. Their findings also show that the disk remains stable for the zero-metallicity case with $\rm Q \geq 1$. However, for the lower metallicities the disk becomes highly unstable. Therefore, masses of the stars in metal polluted disks are expected to be much smaller.

We have assumed here that the disk is in a steady state but it may not always be the case. Our model is unable to capture 3-D effects such as n-body interactions where some of these clumps may get ejected and eventually form stars. We have considered here the two different regimes of optically thick $\rm H_2$ line cooling and $\rm H_2$ CIE cooling. As a result, there is a minor discontinuity at the transition between these regimes. The latter will vanish in a realistic disk where the transition occurs more gradually. It has however no impact on our main results. To enhance our understanding of primordial star formation, we propose to pursue 3D simulations investigating the fragmentation properties of primordial disks at later stages of their evolution to quantify their fragmentation and migration behavior.



\section*{Acknowledgments}
We thank Marta Volonteri, Tilman Hartwig and Kazu Omukai for interesting discussions and helpful suggestions. The research leading to these results has received funding from the European Research Council under the European Community's Seventh Framework Programme (FP7/2007-2013 Grant Agreement no. 614199, project ``BLACK''). DRGS is indebted to the Scuola Normale Superiore (SNS) in Pisa for an invitation as a 'Distinguished Scientist' which has stimulated the research described in this manuscript.

\bibliography{disk.bib}

\end{document}